\documentclass[10pt,a4paper,reqno]{amsart}
\usepackage{amsthm,amssymb,amsmath,graphicx, wrapfig, array, multirow}
\usepackage{mathrsfs,setspace,pstricks, multicol, latexsym}
\newcommand{\toprule}{\hline}
\newcommand{\bottomrule}{\hline}
\newcommand{\midrule}{\hline}

\newtheorem{theorem}{Theorem}[section]
\newtheorem{rem}[theorem]{Remark}

\DeclareMathOperator*{\argminA}{argmin}

\usepackage{listings}
\usepackage{graphicx}
\usepackage{epsfig}
\usepackage{hyperref}
\usepackage{multirow}
\usepackage{wrapfig}
\usepackage{dsfont}
\usepackage{float}
\hypersetup{
	colorlinks,
	citecolor=blue,
	filecolor=black,
	linkcolor=red,
	urlcolor=black
}
\advance\oddsidemargin by -1.6cm
\advance\evensidemargin by -1.6cm

\begin{document}
\author{Anindya Goswami*}
\address{IISER Pune, India}
\email{anindya@iiserpune.ac.in}
\thanks{* Corresponding author}

\author{Sharan Rajani}
\address{Orange Quant Research LLP, Pune, India}
\email{sharan.rajani@orange-quant.com}

\author{Atharva Tanksale}
\address{IISER Pune, India}
\email{atharva.tanksale@students.iiserpune.ac.in}

\title{Data-Driven Option Pricing using Single and Multi-Asset Supervised Learning}

\thanks{This research was supported in part by the SERB MATRICS (MTR/2017/000543), DST FIST (SR/FST/MSI-105), NBHM (02011/1/2019/NBHM(RP)R\&D-II/585).}

\begin{abstract}
We propose three different data-driven approaches for pricing European-style call options using supervised machine-learning algorithms. These approaches yield models that give a range of fair prices instead of a single price point. The performance of the models are tested on two stock market indices: NIFTY$50$ and BANKNIFTY from the Indian equity market. Although neither historical nor implied volatility is used as an input, the results show that the trained models have been able to capture the option pricing mechanism better than or similar to the Black-Scholes formula for all the experiments. Our choice of scale free I/O allows us to train models using combined data of multiple different assets from a financial market. This not only allows the models to achieve far better generalization and predictive capability, but also solves the problem of paucity of data, the primary limitation of using machine learning techniques. We also illustrate the performance of the trained models in the period leading up to the 2020 Stock Market Crash (Jan 2019 to April 2020).\\
\end{abstract}
	
\maketitle
\noindent \textbf{MSC-class:} 91G70, 68T07, 91G20, 91G60\\
\noindent \textbf{Keywords:} Option Pricing, Computational Finance, Learning in Financial Models, Learning and Adaptation
\section{Introduction}
\noindent
Fair pricing of financial instruments is at the heart of market stability. Mispricing securities may cause traders to incur massive losses and can also indirectly affect the financial health of a market. It is thus vital to be able to derive the fair price of tradable financial instruments. The seminal paper \cite{BS} laid the foundation of the theory of no arbitrage option pricing, following which the scope of the theory has been extended by several authors. However, the fair price of an option contract depends on the current anticipation of the future dynamics of its underlying asset. This is why the authors of \cite{HLP} argued that the success or failure of theoretical option pricing and hedging is closely tied to the success in capturing the dynamics of the underlying asset's price movements. Since this is a hard problem, adoption of data-driven approaches in pricing option contracts is gaining attention with the advent of superior computational power and advancements in statistical learning techniques. In this manuscript, we propose data-driven approaches for prescribing the fair price of an option contract without assuming any particular theoretical law of the underlying asset dynamics. We also propose and illustrate the use of data drawn from multiple assets/sources to train these data-driven option pricing models. This allows us a way to mitigate the possible paucity of data available to train models. We believe this might be a promising avenue for further research and could potentially bring some new insights into the field of data-driven option pricing. We would also like to emphasize that the work presented in this study does not attempt to emulate the Black-Scholes formula or any other theoretical option pricing model.\\

\noindent In the past, several authors have investigated the possibility of building a data-driven option pricing model; We give a brief overview of the literature that exists. In \cite{M2S}, the authors conveyed their belief that the trading process of option contracts itself may reveal analytical models. The data-driven investigations in \cite{HLP} and \cite{M2S} were based on option contracts on the S\&P $500$. While the former used only the moneyness parameter (ratio of spot and strike values) and time-to-maturity as inputs to their learning model, the latter also used historical volatility, interest rate, and lagged prices of the underlying asset and option contract. The authors of \cite{QM} obtained a better prediction performance than \cite{HLP} by including the open interest in addition to all the non-lagged inputs of \cite{M2S}. On the other hand, in \cite{M1S}, S\&P $100$ data was used to predict the implied volatility instead of the option price, using past volatilities and option-contract parameters. In \cite{PLJ}, a variant of implied volatility was used as an input to predict the deviation of the actual market price from the Black-Scholes price of the option contract. The model performance was illustrated on AO SPI Index options. If the log returns of the underlying asset is independent of the stock price level, the formula for fair price of an option is homogeneous of degree one in both spot and strike. The authors of \cite{GG} implemented this relation in the structure of the neural network and built a model using option contract data of the S\&P 500 Index. The authors of \cite{CSS} discuss how a technique named profiling could be used to select the optimal neural network structure to predict the implied volatility. This technique was illustrated on USD/NEM exchange rate options and the model took various contract parameters as inputs. The authors of \cite{YLT} argued that option contract data should be partitioned according to moneyness in order to improve the accuracy in pricing options and they illustrated this performance improvement using Nikkei $225$ Index option contracts. In \cite{GQ} the authors exhibited the effectiveness of cross validation, Bayesian regularization, early stopping and bagging in preventing overfitting and improving generalization, in the process of pricing  S\&P $500$ call options using an artificial neural network (ANN). The author of \cite{Ami} attempted to predict the bid-ask spread of options on the OMX Stockholm $30$ Index, using multiple lagged asset prices and their sample standard deviations. In \cite{BS2}, the authors used the dividend rate in addition to Black-Scholes-based features to price options contracts on the FTSE $100$ Index; the model performance was compared with the Black-Scholes-Merton price that incorporates dividends. The authors of \cite{GGK} used S\&P $500$ option contract data and developed a ``modular'' ANN model for option price prediction. In particular, they divided the data set into 9 disjoint parts or modules, according to the moneyness and the time to maturity parameters of the contracts. A similar modularity is adopted in \cite{Das} where the authors build a hybrid model using BANKNIFTY option contracts. Some of the previously mentioned papers have prescribed data-driven option hedging strategies, while some others have also demonstrated success in predicting the price of exotic options using their model outputs. The above survey is not meant to be exhaustive but conveys the broadly accepted methodologies for developing supervised learning models to price options. This paper borrows aspects like homogeneity hint and modularity from the existing literature.\\

\noindent  While a risk-neutral measure is the backbone of theoretical option pricing models, the same is not relevant in an empirical setting. In theoretical models, the No Arbitrage (NA) principle necessitates the usage of a risk-neutral measure. Whereas, in an empirical setting, NA is ensured by the presence of a large number of market participants. In the context of data driven option pricing, one aims to capture the relationship between the historical market data and the traded option price. Therefore, the machine learning algorithm would learn fair pricing only if the historical data does not include instances of arbitrage. Since checking for NA in historical data is a hard problem, we consider only those underlying assets whose option contracts have a high liquidity (in terms of high values of Open Interest). Considering this, we choose to study NIFTY50 and BANKNIFTY index option contracts, the principal indices of the NIFTY exchange. \\

\noindent
In this paper, we propose three different approaches to generate feature sets from the market data, each of which yields $17-22$ features. Each feature set is then used to train two models - using an ANN and the XGBoost algorithm respectively. The underlying assumption throughout the paper is that the statistical distribution of the underlying assets' returns are independent of the level of the stock price ($S$). This implies that the option price function is homogeneous of degree one in both, the spot price ($S$) and the strike price ($K$). In view of this, we construct feature sets using the underlying asset's log returns, moneyness ($\frac{S}{K}$), and time to maturity. Furthermore, the output variable has been constructed using the ratio ($\frac{C}{K} \times 100$) of option price ($C$) to the strike price ($K$). The fair price of an option contract must depend on the anticipated statistical distribution of the future price of the underlying asset. We try to incorporate this principle using a non-parametric approach, wherein we consider a fixed number of consecutive Order Statistics of log returns of the daily underlying close prices as features. We compare the performance of this approach with another approach, wherein the feature set consists of only the first two moments of the log returns of the underlying asset's daily Open-High-Low-Close prices. Both the approaches appear to be similarly effective. Finally we compare these two approaches with a third approach, that augments features from the second approach by including a few additional features derived from the historical option price data. This particular approach outperforms the previous two as the option price data contains significant additional information relevant to the present day option price. To the best of our knowledge, option pricing models using these feature sets have not been reported in the literature so far.\\

\noindent In the proposed data-driven approaches, disjoint consecutive intervals of the option contract price are set as the output instead of a single predicted option price, as we believe that no real market is complete. In other words, a random payoff such as an option contract may have multiple fair prices, and a single predicted price is more confusing than convincing. We therefore define the output variable in a manner that conveys a range of fair prices, in turn formulating option pricing as a bin classification problem rather than a regression problem. We measure and compare the performance of the models described in the manuscript using two different error metrics. The first proposed error metric attempts to mimic the mean absolute error (MAE) while the second metric gives the inaccuracy in predicting the option price to lie within a certain neighborhood of the actual option price. We also compare the performance of the proposed models with the theoretical Black-Scholes option pricing formula with the help of these metrics. The reason behind considering the Black-Scholes model as a benchmark is its straightforward calibration. It is observed that the models constructed using the Approach III outperform the Black-Scholes pricing formula in terms of the above mentioned metrics whereas other proposed models perform equivalently, if not better.  \\

\noindent
We clarify here that, in this paper we do not intend to propose a particular ANN or XGBoost architecture, but rather methodologies through which a broad class of supervised learning algorithms can be used for capturing the option pricing mechanism using market data. Keeping in mind our computational limitations, we select ANN and XGBoost algorithms to illustrate the effectiveness of our methodology. \\

\noindent We would also like to emphasize on the fact that none of the features were selected based on feature importance analysis, as the process of determining feature importance essentially depends on the particular choice of the training data used. Despite maintaining such indifference, the success in predicting option prices indicates that perhaps these data-driven models are capable of learning certain universal rules of option pricing. We also ensure that the inputs and outputs of the models are scale-free, which allows us to investigate if models could be trained on option contract data from two different assets/sources. This, in principle, allows us to construct models that can capture the option pricing mechanism for a broader range of underlying asset dynamics. Our experiments show that the models trained using data from multiple assets/sources possess superior option pricing capabilities than the models trained on individual assets/sources. These experiments have been performed using NIFTY$50$ and BANKNIFTY option price data. However, since we have not experimented with a sufficiently broad class of assets, the complete scope and the limitations of this technique (referred to as combined training) is still unclear. Nevertheless, we propose a methodology to gain a deeper understanding of the combined training effect than what the error metrics offer. This method entails, testing a trained model with family of test datasets derived using simulated Black-Scholes option price data with varying volatility. Results show that the simple idea of combined training results in models that predict the option price for a wider range of underlying asset price dynamics fairly well. In other words we observe domain adaptability for a wide variety of simulation data, clearly indicating the effectiveness of the combined training technique. \\

\noindent Drawing from the modularity approach proposed by \cite{YLT}, \cite{GGK} and \cite{Das}, we choose to train our models on a particular subset of the contract data. We perform our experiments on a ``filtered" dataset comprising of only near-ATM (at-the-money) contracts. The ``filtered" dataset also excludes option contracts that have either too short or too long time-to-maturity values. We believe that including a full range of modularity, as in \cite{YLT}, \cite{GGK}, and \cite{Das}, would complicate the exposition of this paper with too many experiments, as we study six different models constructed using three approaches and two algorithms, on two different assets/sources.\\

\noindent This paper is organized in eight sections. The second section briefly presents the basics of supervised learning, and explains the two supervised learning algorithms used to construct the models. Section \ref{3} contains details about the data under consideration. The input and output of the learning models are explained in Section \ref{4}. In Section \ref{5} we report the performance of the trained models. An analysis of the combined-trained models' performance is presented in Section \ref{6}. Performance of the models on 2019-2020 data is given in Section \ref{7}. Finally we comment on future research directions in the last section.

\section{Supervised Learning Algorithms} \label{2}
\noindent Attempts to develop algorithms that are capable of performing a task without explicitly specifying the expected outcome have led to the development of the field of Machine Learning. This manuscripts leverages a specific subset of machine learning algorithms, known as supervised learning algorithms. These algorithms take in labelled data as input and ``learn" the task at hand. The term ``learn" implies that the algorithms construct abstract representations of the data with the aim of capturing patterns that are fundamental to the task at hand. In the following subsections, we describe briefly two supervised learning algorithms, namely Extreme Gradient Boosting (XGBoost) and Artificial Neural Network (ANN). These algorithms are used in the later sections of this manuscript. Before studying the specifics of the algorithms, it is instructive to understand the general premise of supervised learning algorithms. \\

\noindent
Consider a finite labelled dataset represented as $\{(X_1, Y_1), (X_2, Y_2), (X_3, Y_3), \ldots , (X_J, Y_J)\}$, where the vector $X_j$ is associated with a label $Y_j$. The algorithms attempt to find a mapping $f:X_j \mapsto Y_j$ such that the mapping obtained is the ``best" out of all the possible mappings. A qualitative assessment of the mapping (also referred to as a model) is made possible by an ``objective" function (also known as a ``loss function"). The specifics of the objective function and the strategy used to create the mappings vary with the choice of the algorithm.

\subsection{Extreme Gradient Boosting}
Developed by Tianqi Chen in $2016$ (refer \cite{TC}), Extreme Gradient Boosting combines two powerful techniques, namely ``boosting" and ``gradient descent". It builds upon the gradient boosting decision tree algorithms developed by Friedman in $2001$ (refer \cite{JHF1}) and $2002$ (refer \cite{JHF2}). Gradient boosting involves constructing an ensemble of ``weak" learners, which in the case of XGBoost, are decision trees. These ``weak" learners are combined in an iterative fashion to obtain a ``strong" learner. A ``weak" learner is a model whose accuracy of predictions is slightly better than a model making random predictions. Refer to \cite{FS} for more details on how ``weak" learners can be combined to create ``strong" learners. \\

\noindent
A typical classification task involves categorizing an input to its label (or class). Successfully performing classification requires the model to determine a close approximate of the true conditional probabilities of the classes, given an input. The XGBoost algorithm, for a set of $N$ output classes, assigns a score $F_i(x)$ to the $i^{\text{th}}$ class for the input $x$. We define $F(x)$ as $(F_1(x), F_2(x), F_3(x), \hdots F_N(x))$. The scores obtained are then used to calculate the probability of each class to be the predicted class by using the softmax function, $P(x)$ defined as
\begin{equation}\label{soft}
P_{i}(x) := \frac{e^{F_{i}(x)}}{\sum_{k=1}^{N}e^{F_{k}(x)}}, \quad i = 1, 2, \ldots, N.
\end{equation}

\noindent
The XGBoost algorithm then computes the ``objective" (or loss) function value for each input $x$ by determining how far away from the true distribution is the distribution of the predicted values. This is done by using Categorical Cross Entropy (CE), a loss function, which is defined as
\begin{equation}\label{eq1}
L(z, F(x)) := -\sum_{i=1}^N z_i \; \text{log}(P_i(x))
\end{equation}

\noindent
where $z:=(z_1, z_2, \hdots z_N)$ is a given p.m.f of the true outputs. The XGBoost algorithm seeks to minimize the value of this loss function over all possible $F(x)$ based on the training set of $J$ input-output pairs, $\{(x_j, y_j) \;|\; j = 1, 2, 3, \hdots J\}$. These pairs are used to compute the value of $z^{(j)}$, for each $j$, such that $z_i^{(j)} = \delta(y_j, i)$, where $\delta$ is the Kronecker Delta function. XGBoost then uses the gradient boosting algorithm to obtain an approximate of the minimizer
$\widehat{F}(\cdot) := \; \argminA_{F(.)} \frac{1}{J}\sum_{j=1}^J \; L(z^{(j)}, F(x_j)).$
In order to find the minimizer, the method of steepest descent is applied. The algorithm first computes
\begin{equation*}
    F^{(0)}(\cdot) := \; \argminA_{\gamma \in \mathbb{R}^N} \sum_{j=1}^J L(z^{(j)}, \gamma).
\end{equation*}

\noindent
This iterative scheme is then carried out over $m = 1, 2, 3, \hdots M$ iterations. At $m^{\text{th}}$ iteration the algorithm computes for each $j$, the residual
\begin{equation*}
    r_j^{(m)} := - \frac{\partial L(z^{(j)}, \gamma)}{\partial \gamma} \bigg|_{\gamma \; = \; F^{(m-1)}(x_j)} \; \in \; \mathbb{R}^N.
\end{equation*}

\noindent
The weak learner $h^{(m)}(x)$, is then fit to the training dataset $\{(x_j, \; r_j^{(m)})\}_{j = 1}^{J}$. The algorithm then computes the multiplier $\alpha^{(m)}$ using the equation
\begin{equation*}
    \alpha^{(m)} = \argminA_{\alpha \in \mathbb{R}} \sum_{j=1}^{J} L(z^{(j)}, F^{(m-1)}(x_j) + \alpha h^{(m)}(x_j)).
\end{equation*}

\noindent
This multiplier, $\alpha^{(m)}$, is then used to update the model/score as given by the scheme
\begin{equation*}
    F^{(m)}(\cdot) = F^{(m-1)}(\cdot) + \alpha^{(m)}h^{(m)}(\cdot).
\end{equation*}
\noindent
The XGBoost algorithm thus results in a strong learner by combining $M$ weak learners in order to obtain a close approximate to the true probability distribution. The reader is encouraged to consult the references cited as this exposition is not meant to be comprehensive.

\begin{figure}[!h]
    \centering
    \includegraphics[width = 0.45\textwidth]{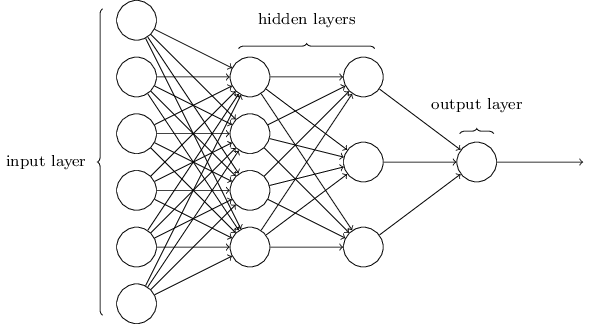}
    \caption{A representative Feed Forward Neural Net}
    \label{fig:general_neural_net}
\end{figure}

\subsection{Artificial Neural Network}
Developments in the field of machine learning led to the advent of algorithms that sought to mimic biological neural networks. These algorithms (referred to as ANN) attempt to harness the ability of biological networks to learn patterns within data. This manuscript presents a brief overview of a special type of ANN known as Feed Forward neural network \footnote{The manuscript uses the term ANN to refer to Feed Forward neural networks in the later sections}. We use Feed Forward neural networks for the experiments proposed in the later sections to classify structured data inputs. The reader may refer to \cite{GYB} for a comprehensive study of ANNs.\\

\begin{figure}[!h]
    \centering
    \includegraphics[width = 0.45\textwidth]{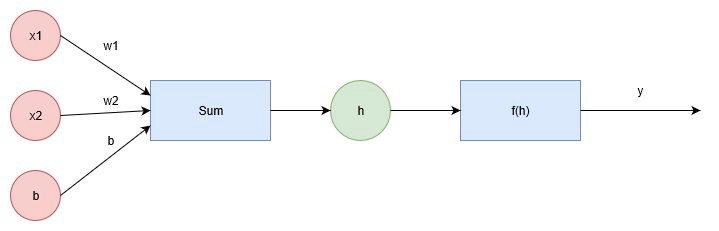}
    \caption{Components of a single neuron}
    \label{fig:neuron_rep}
\end{figure}

\noindent
\footnote{Figure taken from \url{http://neuralnetworksanddeeplearning.com/chap1.html}}Figure \ref{fig:general_neural_net} depicts a general Feed Forward neural network (this figure is representative. Refer Table \ref{tab:neural_net_structure} for details). A neural network is a set of ``neurons" that interact with each other to ``learn" the representation space of the input data. Figure \ref{fig:neuron_rep} shows the structure of a neuron.

\noindent
As can be seen in Figure \ref{fig:neuron_rep}, the output $\eta$ of a neuron can be given by-
\begin{equation*}
    \eta = f\bigg(\sum_{i=1}^{n} w_i\psi_i + b\bigg)
\end{equation*}
where $\psi = (\psi_1, \; \psi_2, \hdots,\; \psi_n) $ are the inputs to the neuron, $w_i$ is the weight associated with each input $\psi_i$ and $b$ is the overall bias associated with the neuron; the function $f$ is called the activation function and is used to impart non-linearity to the neural network. As evident from Figure \ref{fig:general_neural_net}, a feed forward neural network consists of a number of ``layers" of stacked neurons. Each neuron in a layer is connected to every neuron in the next layer. Thus the outputs of the neurons in the preceding layer act as the inputs to the neurons in the next layer. As stated earlier, each ``connection" between any pair of neurons, has a weight $w$ associated with it. The optimal number of layers in a neural network and the number of neurons in each layer is to be determined for a given problem, and is referred to as the architecture of the neural network. Along with this, it is also necessary to determine the appropriate activation functions for each of the neurons as well as the optimization scheme to be used. The architecture of the ANN used in the present study (subject to our computational constraints) is given in Table \ref{tab:neural_net_structure}.

\begin{table}[!h]
\centering
\begin{tabular}{@{}ccc@{}}
\toprule
\multicolumn{3}{c}{\textbf{Composition of the ANN used}}                     \\ \toprule
                 & \textit{Number of Neurons} & \textit{Activation Function} \\ \midrule
\textit{Layer 1} & 128                        & ReLU                         \\
\textit{Layer 2} & 64                         & ReLU                         \\
\textit{Layer 3} & 50                         & softmax                      \\ \bottomrule
\end{tabular}
\caption{``Architecture" of the Neural Net used}
\label{tab:neural_net_structure}
\end{table}

\noindent
The activation function used for each layer has been indicated in Table \ref{tab:neural_net_structure}. The ReLU activation function is defined as
$$ \text{ReLU}(x) = \text{max}(0, x). $$

\noindent
The softmax function, as explained previously (refer Equation (\ref{soft})), gives the class probabilities. We use the loss function, categorical crossentropy (refer Equation (\ref{eq1})) to determine how far the true probability distribution is from the distribution of the predicted values. In order to ``learn" a given task, the sequence of weights that serve as a minimizer to the loss function are to be found, as this corresponds to a higher prediction accuracy by the neural network. This is achieved by optimizing the weights using an optimization scheme (commonly known as training the network). In the present study, we use the Adam optimiser, an advancement of the stochastic gradient descent optimizer (refer \cite{KB}). \\


\section{Data}\label{3}
\noindent NSE, an Indian stock exchange, facilitates the trading of option derivatives on stocks and stock indices in high volumes. Markets with high trading volumes generally imply a high level of trader participation, which further implies a lower chance of the market being imperfect (i.e, the market is efficient). This also allows us to consider the traded price of the derivative as the ``fair" price. Persistent high trading volumes for a particular range of option contracts give us a better chance to ``learn" the pricing mechanism of those option contracts. Some of the NSE based stock indices that have a high option contract trade volumes are the NIFTY$50$ and BANKNIFTY. For our experimentation, we extract the daily contract price data of European call options for both, NIFTY$50$ and BANKNIFTY. Data is obtained for the years $2014-2018$, from the NSE website's contract wise archive section\footnote{The data can be accessed using the link - \url{https://www1.nseindia.com/products/content/derivatives/equities/historical_fo.htm}}.
 \begin{figure}[!h]
     \centering
     \includegraphics[width = 0.9\textwidth]{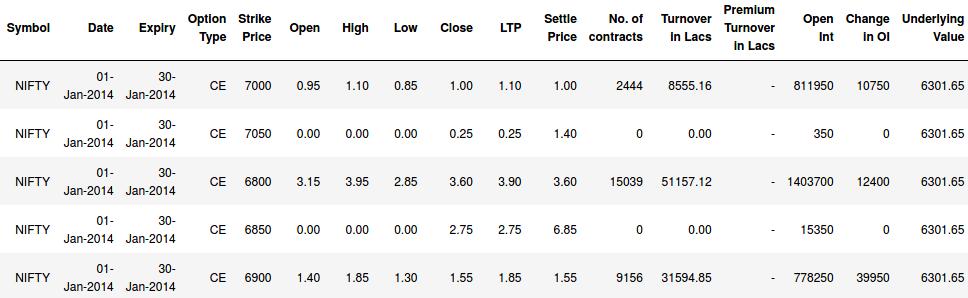}
     \caption{Snapshot of the Unfiltered Option Dataset}
     \label{fig:option_data}
 \end{figure}
It is then ensured that the data set obtained is purged of contracts that are not traded. For reasons related to the construction of the models, we add a new column to the filtered dataset that records the close price of the same option on the previous day. If the option contract did not exist on the previous day, we report the value $0$ in this new column. We subsequently screen the data to remove all rows that have a zero in the new column. We then add more columns to the data array to include the ``Open", ``High", ``Low" and ``Close" prices of the underlying asset for the past 20 days corresponding to each row. Further more, we add an additional column that gives us the three months' government bond yield rate (see Section \ref{4.2}). \\

\noindent We then select the option contracts that are in the vicinity of at-the-money(ATM) contracts, i.e., we only select those contracts for which the quantity $|1 - \frac{S}{K}|$ is not more than the pre-decided value of $0.04$, where $K$ and $S$ are the strike and the spot prices respectively. We refer to such contracts as near-ATM option contracts. It has been observed that numerous near-ATM option contracts are traded everyday with identical or different time to maturities. However, significantly low trading volume is observed for contracts with very large or very small time to maturities. Hence we choose to study, only those contracts whose time-to-maturity values are not more than 45 days and not less than 3 days.\\

\begin{table}[ht]
\begin{tabular}{@{}ccc@{}}
\toprule
\textbf{\textit{Dataset}}      & \textbf{NIFTY50} & \textbf{BANKNIFTY} \\ \midrule
\textit{Raw}      & 1072695    & 350186               \\ \midrule
\textit{Filtered} & 13516      & 20414                \\ \midrule
\textit{Train}    & 10837      & 13622               \\ \midrule
\textit{Test}     & 2679       & 6792               \\ \bottomrule
\end{tabular}
\caption{Train/Test Split: Dataset Sizes}
\label{tab:test_train_split}
\end{table}

\noindent \textbf{Train/Test Split} Figure \ref{fig:option_data} is an indicative sample of the NIFTY$50$ option contract dataset that we obtain from the NSE. In order to build a predictive model using the algorithms described in Section \ref{2}, the dataset needs to be split into separate datasets that would be used to train and evaluate the trained models. Most supervised learning algorithms when trained with time series data, necessitate splitting the dataset linearly as individual observations are not independent. In the same vein, we split the dataset into approximately two linear parts in the ratio $70:30$. The $33$ months period, i.e., data from Jan $2015$ to Sept $2017$ forms the training dataset and the succeeding data i.e., from Oct $2017$ to Dec $2018$ forms the test dataset for evaluating the proposed models. Table \ref{tab:test_train_split} shows the number of datapoints we deal with at every step of the model building and evaluation process.

\section{Model I/O}\label{4}
\noindent
As mentioned previously, this study aims to develop supervised learning models that can ``learn" the market perceived pricing of option contracts, and give us the fair price of an option contract in accordance with past market behaviour. In order to develop supervised machine learning models (refer Section \ref{2}), we need to train the models with a set of `inputs" and ``outputs". Sections \ref{4.2}, \ref{4.3} and \ref{4.4} describe the different feature sets, each of which we intend to use as inputs to the supervised learning algorithms. Other than the homogeneity hint, we make no other assumption with regards to asset dynamics. On the other hand, if we include feature variables that estimate parameters of a parametric model for asset dynamics, it will essentially create a model that learns the option price equation under that parametric model. Since we wish to investigate option pricing in a truly non-parametric setting, we exclude any such estimators (including volatility estimators) from the feature sets. The proposed feature sets are derived from the information available to market participants. Before describing each of the feature sets, we explain the desired format of the output variable which is kept uniform across all the approaches.
\subsection{Categorical Output Variable}\label{4.1} For the output of the proposed data-driven option pricing models, using the option contract prices obtained directly from the market would not be prudent. This is because, for contracts with a fixed value of moneyness, the magnitude of contract parameters like ``Strike" and ``Spot" prices may vary over the years. It makes much more sense to create an output variable that is scale free. We therefore define the ``output" as the ratio---expressed in percentage---of the Close price ($C$) and the Strike price ($K$) of the contract, i.e. we designate $100 \times \frac{C}{K}$ as the output variable. This ratio serves as a scale free proxy to contract price for the model. \\

\begin{figure}[!h]
    \centering
    \includegraphics[width = 0.55\textwidth]{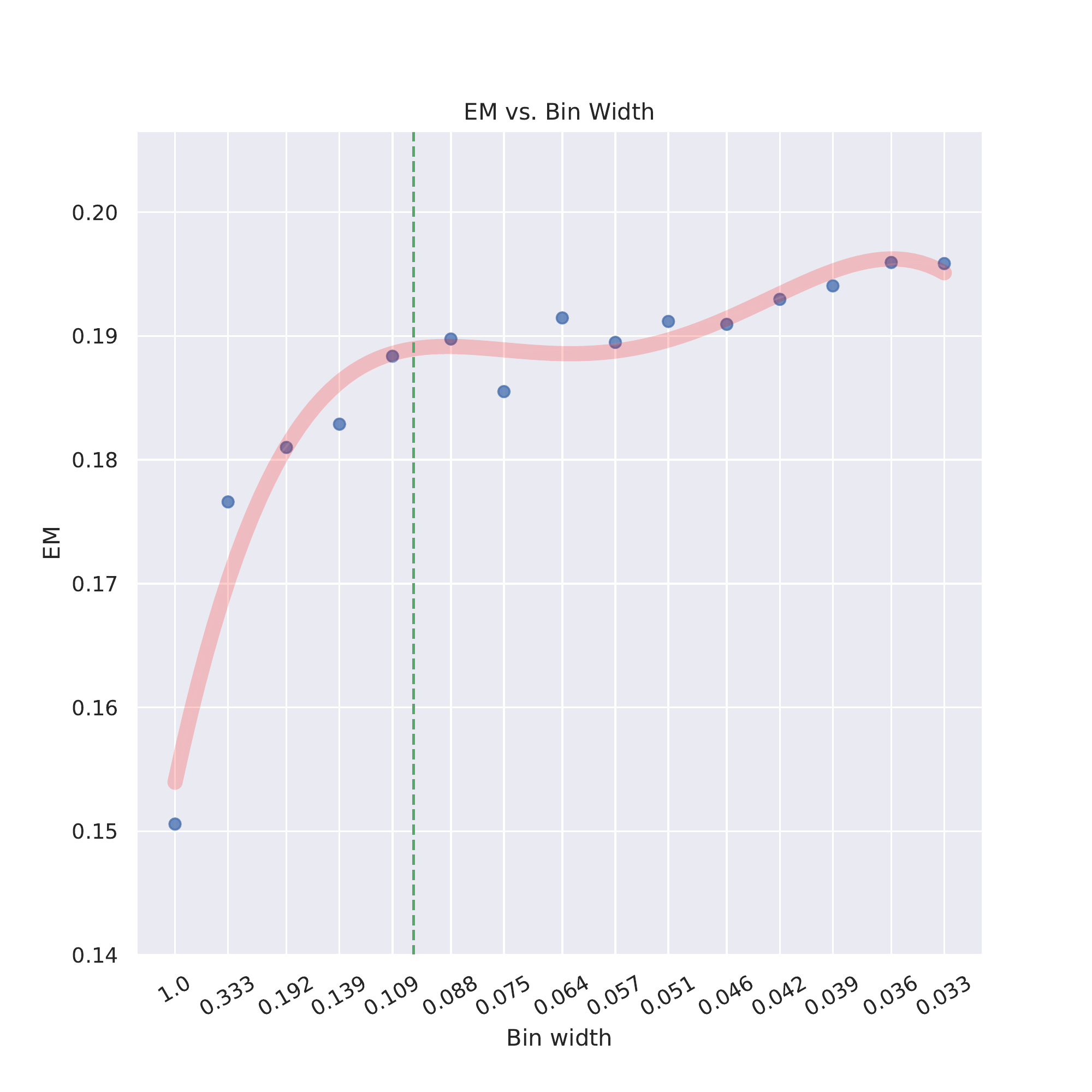}
    \caption{EM vs bin-width for the NIFTY50 dataset}
    \label{fig:EMbin}
\end{figure}

\begin{figure}[!h]
    \centering
    \includegraphics[width = 0.55\textwidth]{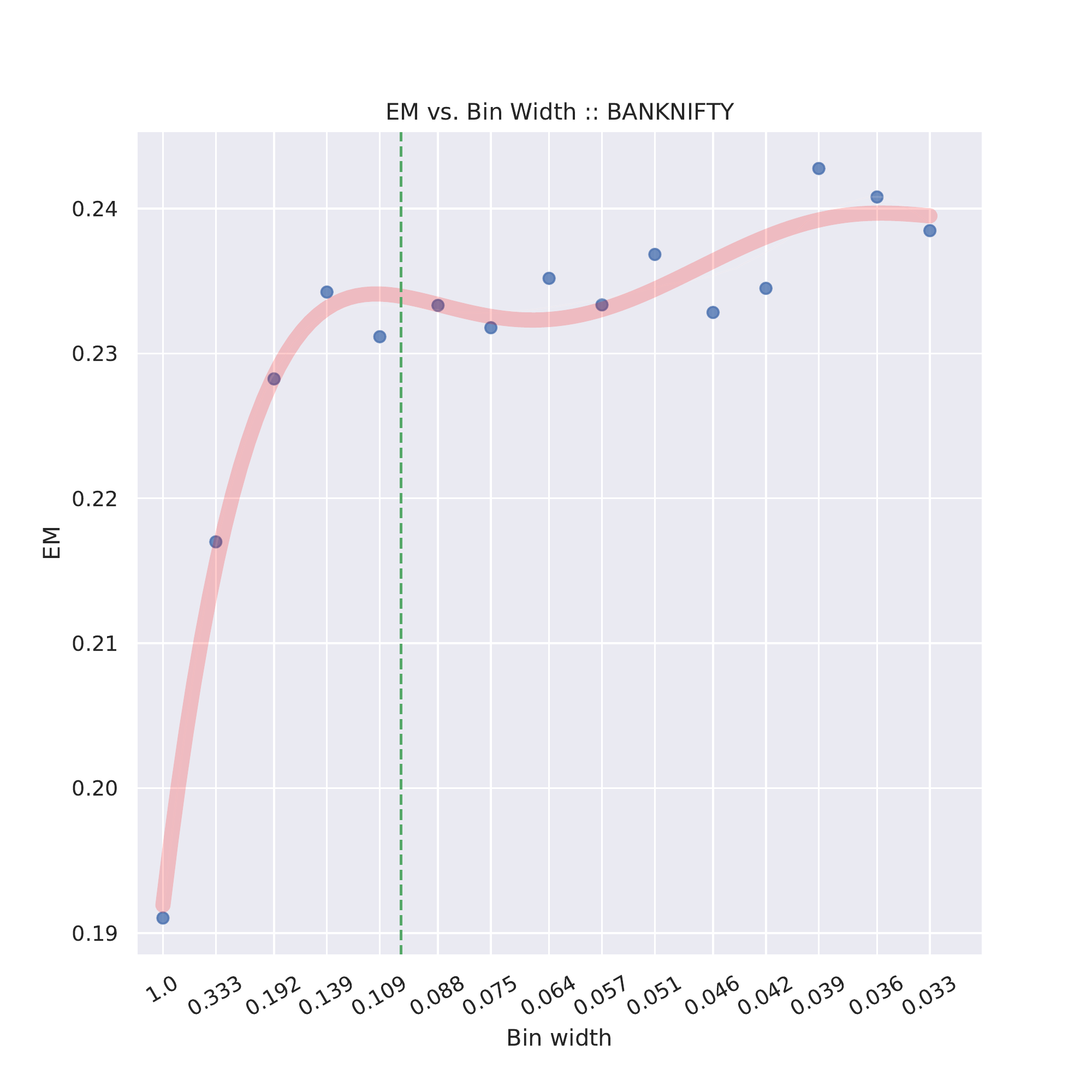}
    \caption{EM vs bin-width for the BANKNIFTY dataset}
    \label{fig:BNF_EMbin}
\end{figure}

\noindent
Since the ``output" variable is continuous, it is natural to formulate the problem using a regression model. However, since no real market is complete, the option contracts may possess multiple fair prices. In view of this a single predicted price is not justified unless an additional optimization on hedging strategies is performed. Without such trader specific assumptions, the fair price could be anything in a certain interval. Determining this interval is a hard problem from both, the theoretical and the empirical aspects. Instead of finding such an interval of the fair prices, selecting the most likely interval from a pre-determined set of non-overlapping consecutive intervals is fairly straight forward. One can divide the range of outputs into non-overlapping ``bins" and select the ``embracing" bin as the output variable.  However, a major hurdle in this approach is determining the width of the bins. Larger the width of each bin, lesser the usefulness of the model due to lack of precision. On the other hand a finer binning confuses the model due to the presence of a certain degree of in-docile uncertainties in the option trading price, which can be attributed to the lack of completeness in the market. To tackle this issue we introduce a binning insensitive performance measure for the models. We refer to this measure as the EM (refer subsection \ref{5.1}, Equation \eqref{EM}). We then study the values of EM obtained for different bin widths, for each of the data sets, for a fixed model type. Since EM is asymptotically insensitive to binning as bin-width is progressively reduced, we choose the largest bin-width for which the EM stabilizes. \\

\noindent
Figures \ref{fig:EMbin} and \ref{fig:BNF_EMbin} show the relation between EM and bin-width obtained for NIFTY50 and BANKNIFTY data-sets respectively. We observe that EM drastically decreases if bin-width increases above $0.1$. This is expected as a larger bin-width implies lesser number of bins, which results in greater accuracy despite poorer precision. On the other hand, for a decrease in bin-width below $0.075$, a certain increase in EM appears, due to higher number of bins. But for bin interval width roughly between $0.1$ and $0.075$, the EM value behaves insensitive to binning in both the figures. Hence we take bin width as $0.1$ and partition the entire range of output values in the manner explained in the following paragraph.\\
\begin{figure}[!h]
    \centering
    \includegraphics[width = 0.55\textwidth]{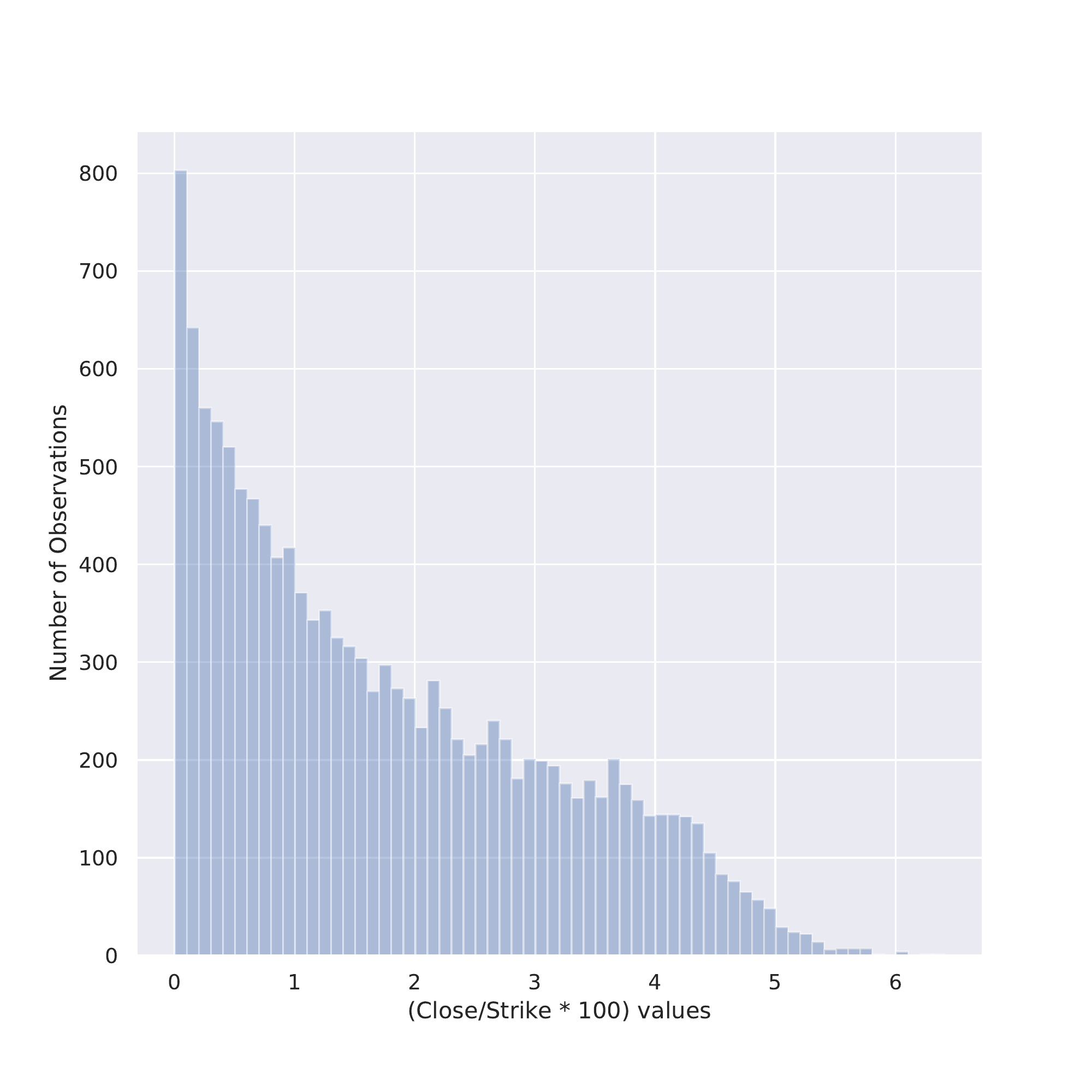}
    \caption{$100\times\frac{C}{K}$ values for NIFTY50 contracts plotted as a histogram}
    \label{fig:percentage_hist}
\end{figure}

\noindent
The interval $((n-1)w, \; nw]$ is set as the $n^{\text{th}}$ bin where $n$ is a natural number and $w$ (here $w = 0.1$) is the bin interval width. This creates a set of equispaced bins allowing us to map option prices $C$ to their respective bins (the bin that includes the value of $100 \times \frac{C}{K}$) and assigning the corresponding integer valued bin number to it as its label. These labels are then considered as the ordinal output variables and are used to train and test the constructed models. We illustrate this binning in Figure \ref{fig:percentage_hist}. The figure is a histogram (plotted using $0.1$ as the bin width) of $100 \times \frac{C}{K} $ values, for the filtered NIFTY$50$ contract dataset. It is evident from the plot that there are just enough data points per bin and yet we have enough number of categories, ie. bins, to make the model robust. \\

\noindent
Since binning defines the output variable in the training and test data-sets for each model, the choice of a fixed bin width, makes models trained on either of the data-sets comparable. Without uniform binning, it would not be possible to combine or compare models trained on different datasets.

\begin{rem}
The following subsections (\ref{4.2}, \ref{4.3} and \ref{4.4}) describe three separate, independent ``approaches" used to generate feature sets that serve as inputs to the supervised learning algorithms described in Section \ref{2}. Here the term ``approach" is used to convey the motivation/idea behind generating the feature sets. The following subsections represent a crucial part of the present study. The components that make up each of the feature sets are summarised at the end of this section in Table \ref{tab:feature_summary}.
\end{rem}

\subsection{\textbf{Approach I}} \label{4.2}
From the very definition of an option contract, it is known that the fair price of a European call option must depend on the values of the option contract parameters (like the strike price ($K$), the time to maturity ($\tau$)), the risk free interest rate ($r$), the spot price ($S$) and the anticipated statistical behavior of the future dynamics of the underlying asset. The closest real world approximate of the value of $r$ would be the government bond yield. Moreover, the future dynamics of the underlying security can be anticipated from its present price and the price dynamics followed by it over the past few days. Directly using a string of historical asset price data as features would make the values scale dependent, especially so when data over many years is to be considered for model training. As a means to resolve the scale dependency, log returns of the time series (henceforth, referred to as LR) are considered, the values of which are given by
\begin{equation}
    \text{LR}(S)_i = \log(S_i) - \log(S_{i-1}) = \log{\frac{S_i}{S_{i-1}}}
\end{equation}
where, $S_i$ is the $i^{th}$ term of a time series $S$. 
 For the sake tractability it is natural to assume stationarity of the log returns. In order to obtain a non-parametric non-temporal inference of the recent distribution of log returns, we calculate the Order Statistics of the log returns. This is done by computing the log returns of the daily close prices of the underlying asset for a window of the past 20 trading days, as it corresponds to approximately a calendar month excluding all holidays. Following this, the Order Statistics is computed by simply arranging the log returns in ascending order for each sample. To be more precise, if $x_{(i)}$ denotes the $i^{\text{th}}$ order statistic of a sample of different real values $(x_1, x_2, x_3, \ldots, x_n)$, then $x_{(i)}=x_j$ for some $j=1, \ldots, n$, where $x_{(1)} < x_{(2)} < \cdots < x_{(n)}$ holds. Hence, considering Order Statistics does not result in any loss of statistical information for samples drawn from a fixed population.\\

\noindent In view of the preceding discussions, we calculate the Order Statistics of historical log returns for each of the near-ATM option contracts resulting in a row of $22$ features as given below:
\begin{enumerate}
\item The $19$ log return order statistics.
\item The time to maturity ($\tau$) of the option contract.
\item The interest rate $r$: We use the $3$ month sovereign bond yield rates as an approximation for the risk free interest rates.
\item Moneyness: Computed as $\frac{S}{K}$ (the ratio of Spot to Strike prices).
\end{enumerate}

\noindent
A collection of such rows is what constitutes the train/test dataset.

\subsection{\textbf{Approach II}}\label{4.3}
\begin{figure}[!h]
    \centering
    \includegraphics[width = 0.3\textwidth]{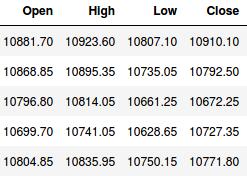}
    \caption{Cross section of the underlying asset price dataset}
    \label{fig:asset_ohlc}
\end{figure}

This subsection proposes a feature set that takes into account the market participant's access to other facets of the asset price data. Intuitively, a lot more information on asset dynamics can be gleaned by taking into account the values of ``Open", ``High", and ``Low" along with the values of ``Close" (refer \ref{fig:asset_ohlc}). However, this intuitive anticipation deserves a quantitative backing.
Let us first understand the need for a completely new ``approach". The previous subsection attempted to generate a feature set that captures the empirical distribution of the ``Close" price data of the underlying asset. The present subsection seeks to remedy the fact that the asset price data obtained from the market consists of multiple facets that haven't been accounted for in Approach I. The joint distribution of these four time series' (``Open", ``High",``Low" and ``Close" ) cannot be inferred from the order statistics of every individual time series as they are not independent. This renders a direct mimicking of Approach I ineffective. Moreover, using a direct extension of Approach I would lead to a feature set with $19 \times 4 = 76$ features. This bloating up of the feature set prevents any meaningful comparison between different models. It is therefore prudent to adopt a moments based approach to generate a feature set that is sensitive to all facets of the underlying asset data. \\

\noindent Instead of trying to obtain an empirical distribution of the multivariate time series, we measure the central tendency and the dispersion using the first raw moment and the covariance matrix of the component-wise log returns of the vector valued series. The feature set is then built using these statistics. Formally put, for a window of the past $20$ trading days, we compute the arithmetic mean of log returns of Open ($O$), High ($H$), Low ($L$), and Close ($C$), denoted by $\mu_O$, $\mu_H$, $\mu_L$, and $\mu_C$ respectively and construct the covariance matrix $\Sigma$ as
$$ \Sigma =
\begin{bmatrix}
    \text{var}({O})       & \text{cov}({O,H}) & \text{cov}({O,L}) &  \text{cov}({O,C}) \\
    \text{cov}({H,O})       & \text{var}({H}) & cov({H,L}) &  \text{cov}({H,C}) \\
    \text{cov}({L,O})       & \text{cov}({L,H}) & \text{var}({L}) &  \text{cov}({L,C}) \\
    \text{cov}({C,O})       & \text{cov}({C,H}) & \text{cov}({C,L}) & \text{var}({C})
\end{bmatrix}.
$$

\noindent As $\Sigma$ is symmetric, six entries on the upper triangular part are repeated in the lower part. We include the square root of entries of $\Sigma$ in the feature set after discarding the repetitions. Thus we build the second feature set using the following 17 features:
\begin{enumerate}
    \item Means of the log return series'; $\mu_O$, $\mu_H$, $\mu_L$, and $\mu_C$.
    \item Ten statistics from $\Sigma$, namely $\left\{\frac{\Sigma_{ij}}{\sqrt{|\Sigma_{ij}|}} \mid 1\le j\le i\le 4\right\}$, where $\Sigma_{ij} $ is the $(i,j)^{\text{th}}$ element of $\Sigma$ using the convention $\frac{x}{\sqrt{ |x|}}=0$ iff $x=0$.
    \item Features $(2) - (4)$ from Approach I.
\end{enumerate}

\subsection{\textbf{Approach III}} \label{4.4}
\begin{figure}[!h]
    \centering
    \includegraphics[width = 0.85\textwidth]{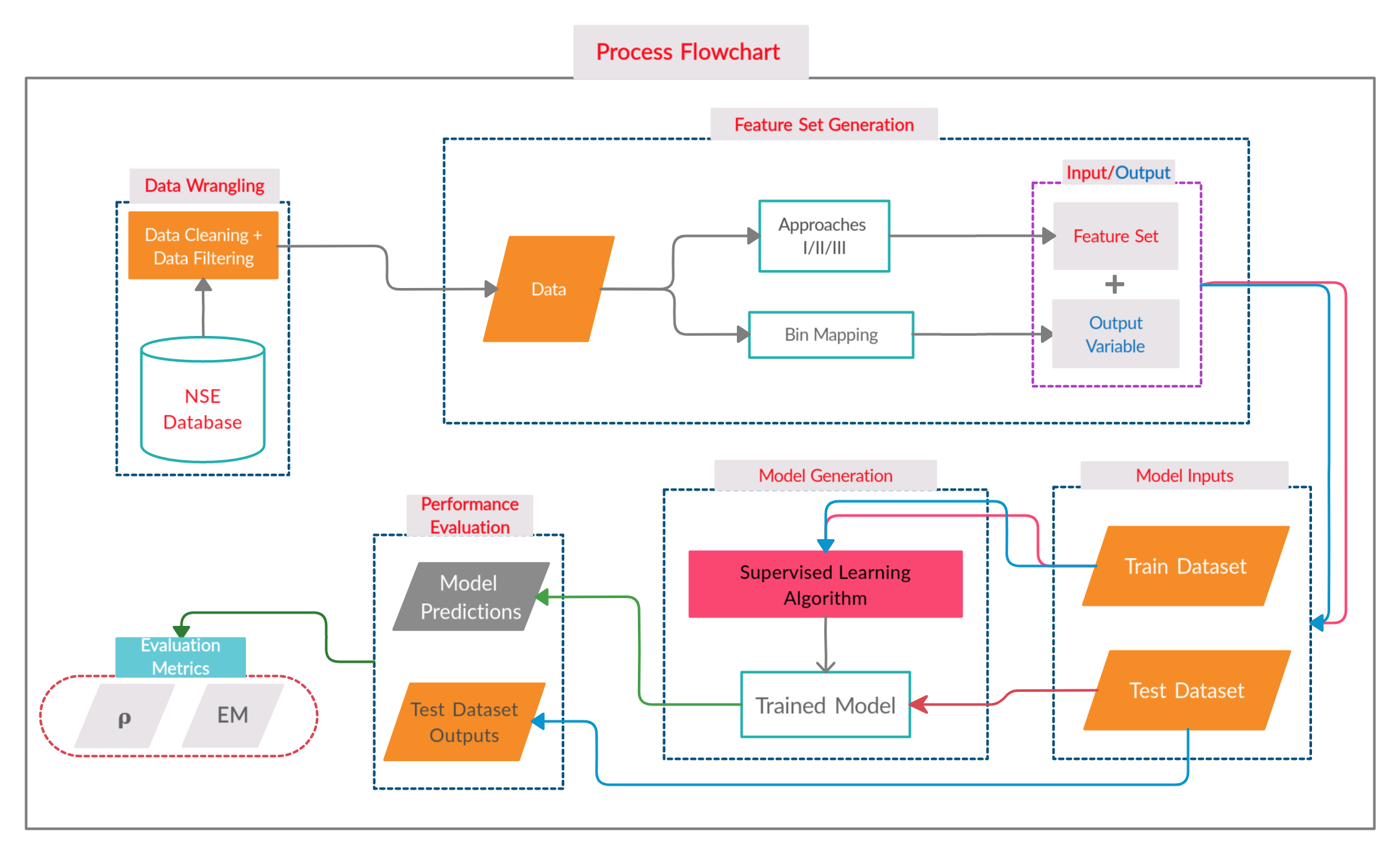}
    \caption{Process Flowchart}
    \label{fig:flowchart}
\end{figure}

Approaches I and II primarily utilize the underlying asset price data to derive the set of features. However, a market participant also has access to the historical option contract trade prices. It would be imprudent to not develop an approach that factors in this key aspect. In fact, including the historical option contract trade prices in an appropriate form would help the supervised learning algorithms to develop abstract representations of market factors like implied volatility, allowing them to predict the option contract price more accurately. We would like to stress on the fact that the intent of Approach III is to build upon the progress made in Approach I and II. We cannot use an extension of Approach I for reasons mentioned previously. We instead, seek to augment the feature set developed in Approach II by adding the features listed below to the feature set obtained from Approach II:
\begin{enumerate}
    \item Previous Option Price (scaled): This is computed as $\frac{C_{t-1}}{K}$ where $C_{t-1}$ is the previously reported close price of the option contract under study and $K$ is the Strike price of the contract. Including this feature helps account for any auto-regressive characteristics that might be present in the option price data.
    \item Mean Moneyness: Computed as $\frac{\bar{S}}{K}$, where $\bar{S}$ is the mean of the underlying asset prices (for a window of the past $20$ trading days) and $K$ is the strike price of the contract.\\
\end{enumerate}
Table \ref{tab:feature_summary} summarizes the features used by the three approaches described in this section. Figure \ref{fig:flowchart} presents an overview of the steps that constitute the process of model building.

\begin{table}[h]
\centering
\hspace*{-1.8cm}
\begin{tabular}{@{}cccc@{}}
\toprule
\multicolumn{4}{c}{\textit{\textbf{Composition of Feature Sets: An overview}}} \\ \toprule
 &
  \textit{Approach 1} &
  \textit{Approach 2} &
  \textit{Approach 3} \\ \midrule
\begin{tabular}[c]{c}\textit{\textbf{Non Parametric}}\\ \textit{\textbf{Features}}\end{tabular}&
  \begin{tabular}[c]{@{}c@{}}Order Statistics \\ of the LR of the\\  underlying asset\end{tabular} &
  --- &
  --- \\
 \multicolumn{1}{l}{} &
 \multicolumn{1}{l}{} &
 \multicolumn{1}{l}{} &
 \multicolumn{1}{l}{} \\
\textit{\textbf{\begin{tabular}[c]{@{}c@{}}Parametric \\ Features\end{tabular}}} &
  --- &
  \begin{tabular}[c]{@{}c@{}} Mean LR of OHLC\\ Cov LR of OHLC\end{tabular} &
 \begin{tabular}[c]{@{}c@{}} Mean LR of OHLC\\ Cov LR of OHLC\end{tabular} \\
\multicolumn{1}{l}{} &
  \multicolumn{1}{l}{} &
  \multicolumn{1}{l}{} &
  \multicolumn{1}{l}{} \\
\textit{\textbf{\begin{tabular}[c]{@{}c@{}}Contract \\ Features\end{tabular}}}
& \begin{tabular}[c]{@{}c@{}}Moneyness\\ Time to Maturity\end{tabular} &
  \begin{tabular}[c]{@{}c@{}}Moneyness\\ Time to Maturity\end{tabular} &
  \begin{tabular}[c]{@{}c@{}} Moneyness\\ Time to Maturity\end{tabular} \\
& & &  \begin{tabular}[c]{@{}c@{}} Prev. Option Price (scaled)\\ Mean Moneyness\end{tabular}\\
\textit{\textbf{Other}}  & Interest Rate & Interest Rate & Interest Rate\\ \toprule
\textbf{Total} &
  \textit{$19 + 3 = 22$ features} &
  \textit{$4+ 10 + 3 = 17$ features} &
  \textit{$4 + 10 + 5 = 19$ features} \\ \bottomrule
\end{tabular}
\hspace*{-1.8cm}
\caption{An overview of feature sets for all the Approaches}
\label{tab:feature_summary}
\end{table}

\section{Model Performance}\label{5}
\subsection{Performance Measures}\label{5.1}
Once a model is trained, it is imperative to test its performance on data that has not been used for training (i.e. the test dataset) and study the quality of the predictions. The most common way to evaluate the predictions of nominal variables is to find the value of the accuracy metric $A$, defined as
\begin{equation}
    A = \frac{C}{T}
\end{equation}
where $C$ is the number of correct predictions and $T$ is the total number of predictions. It is however, not ideal to use the accuracy metric for an ordinal output variable having a wide range. In such cases, one can examine the quality of the incorrect predictions by measuring the distance between the actual and the predicted classes. Doing so is meaningful because, it is desirable for a good model to be able to predict a class identical to or very close to the actual class. In contrast, the accuracy metric treats all incorrect predictions in the same manner, regardless of whether the predicted class is close to or far from the actual class. It is therefore important to come up with a metric that does a better job of informing us about the quality of the predictions made. We do so by proposing an \textit{Error Metric} (\textit{EM}) given by
\begin{equation}\label{EM}
    \text{EM}  = \left( \frac{w}{T}\sum_{i=1}^{i=T}|C_i - P_i| \right)  
\end{equation}

\begin{figure}[!h]
    \centering
    \includegraphics[width = 0.45\textwidth]{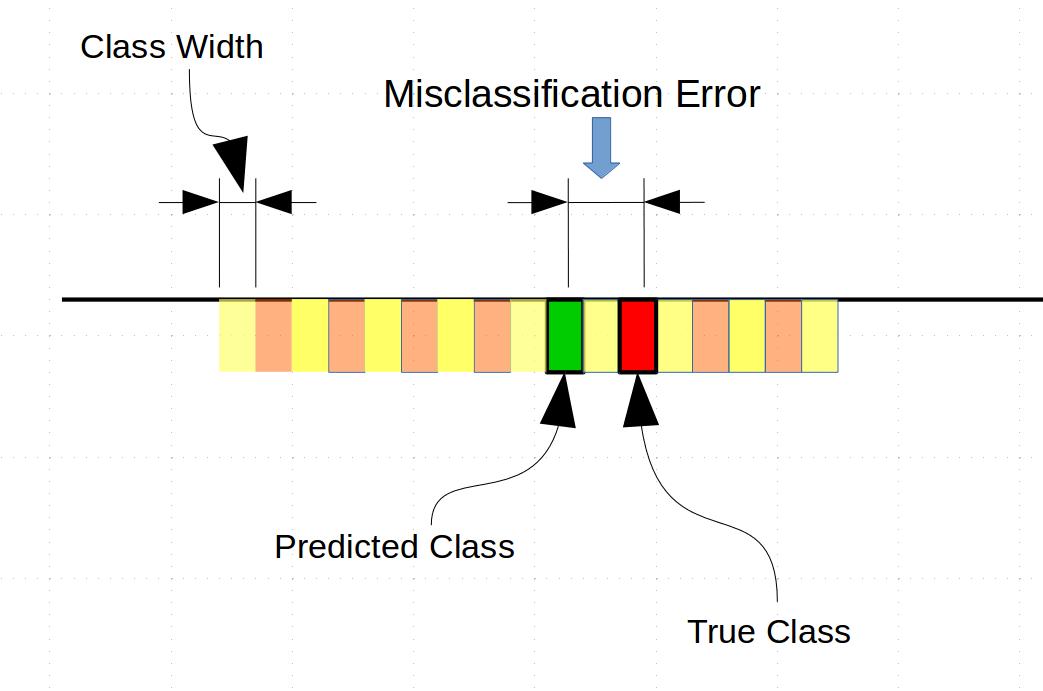}
    \caption{Visualizing the EM for a single prediction}
    \label{fig:nem_visualization}
\end{figure}
\begin{figure}[!h]
    \centering
    \includegraphics[width = 0.45\textwidth]{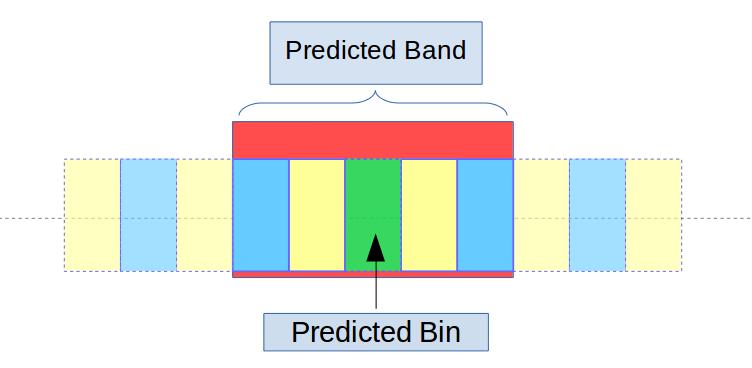}
    \caption{Visualizing the predicted bin, the predicted band and the relationship between them}
    \label{fig:rho_band}
\end{figure}


\noindent
where, $w$ denotes the \textit{binwidth}, $T$  is the number of contracts in the \textit{test} dataset and the ordinal variables $C_i$  and $P_i$ denote the actual and the model predicted bin numbers respectively. As mentioned in Subsection \ref{4.1}, we set the value of $w$ as $0.1$. Multiplying the bin number with the bin width makes EM asymptotically insensitive to binning. We illustrate the implication of EM in Figure \ref{fig:nem_visualization}. Figure \ref{fig:nem_visualization} gives an example of the case where the distance between the actual and the predicted classes is $2$. It can easily be proved that the EM converges to the Mean Absolute Error (MAE) as bin width tends to $0$. However, the MAE metric is known to be sensitive to outliers. Hence, in order to get a better insight into the performance of the models, we also consider an additional metric---the ``inaccuracy metric"---that is robust to outliers. The ``inaccuracy metric" ($\rho$) gives the probability of the predicted and actual bins to lie more than $2$ bins apart. In other words, the metric $\rho$ gives the probability that the model will fail to include the actual price bin (labelled as $C_i$) in a band of five consecutive bins where the predicted bin (labelled as $P_i$) is in the middle. We refer to the above mentioned band as the predicted band (see Figure \ref{fig:rho_band}). The $\rho$ metric is defined as

\begin{equation}\label{rho}
     \rho := \frac{\# \{i\in \{1,2,\ldots,T\} \mid |C_i-P_i| > 2 \}}{T}.
\end{equation}
 While EM is a measure of prediction imprecision, the empirical quantiles of the error $C_i-P_i$ gives the confidence interval of $C_i$ using prediction $P_i$. In particular, $1 - \rho$ denotes the confidence of $C_i$ being in $[P_i-2, P_i+2]$.
\noindent

\subsection{Models Trained with Single Sources}\label{5.2} The following paragraphs present the details of the performance of each model (in terms of EM and $\rho$) built using the aforementioned approaches for the NIFTY$50$ and BANKNIFTY datasets. \\

\paragraph{\textbf{NIFTY$50$ Index option data}}
Table \ref{tab:Main_Reuslts} lists the EM and $\rho$ values for all models that are trained and tested using NIFTY$50$ data.
\begin{table}[!h]
\centering
\begin{tabular}{@{}ccccc@{}}
\toprule
\multicolumn{5}{c}{\textit{\textbf{NIFTY50 Contracts}}}                        \\ \toprule
& \multicolumn{2}{c}{\multirow{2}{*}{\textbf{EM}}} & \multicolumn{2}{c}{\multirow{2}{*}{\textbf{$\rho$}}} \\
 & \multicolumn{2}{c}{} & \multicolumn{2}{c}{} \\ \cline{2-5}
B-S Pricing &  \multicolumn{2}{c}{0.19} & \multicolumn{2}{c}{0.29}\\ \cline{2-5}
\textit{\textbf{Trained Models}}  & \textit{\textbf{ANN}} & \textit{\textbf{XGB}} & \textit{\textbf{ANN}}      & \textbf{XGB}     \\ \midrule
Approach I   & 0.18     & 0.18   & 0.25    &  0.27     \\
Approach II  & 0.17     & 0.19     & 0.26     &  0.29    \\
Approach III & 0.14     & 0.16     & 0.20    &  0.22    \\ \bottomrule
\end{tabular}
\caption{Model evaluation metrics for models trained and tested on NIFTY50 options contract price data}
\label{tab:Main_Reuslts}
\end{table}
The results reported in Table \ref{tab:Main_Reuslts} convey that all trained models perform at par or better than the pricing formula of the Black-Scholes model (we use the historical volatility values observed over a window of the past $20$ trading days to compute the Black-Scholes price). We also note that in comparison to XGBoost, the use of ANN results in lower values of EM and $\rho$. Table \ref{tab:Main_Reuslts} also shows that the values of the metrics do not differ significantly between Approach I and Approach II. This indicates that the two supervised learning algorithms were unable to extract additional information on the asset dynamics from the first two moments of the Open-High-Low-Close (OHLC) data than solely from the Close price data. From the results, it is also clear that the performance of Approach III is far superior to that of Approaches I and II, which indicates that the historical option price data contains valuable information relevant to the current option price.\\

\begin{figure}[!h]
    \centering
    \includegraphics[width = 0.55\textwidth]{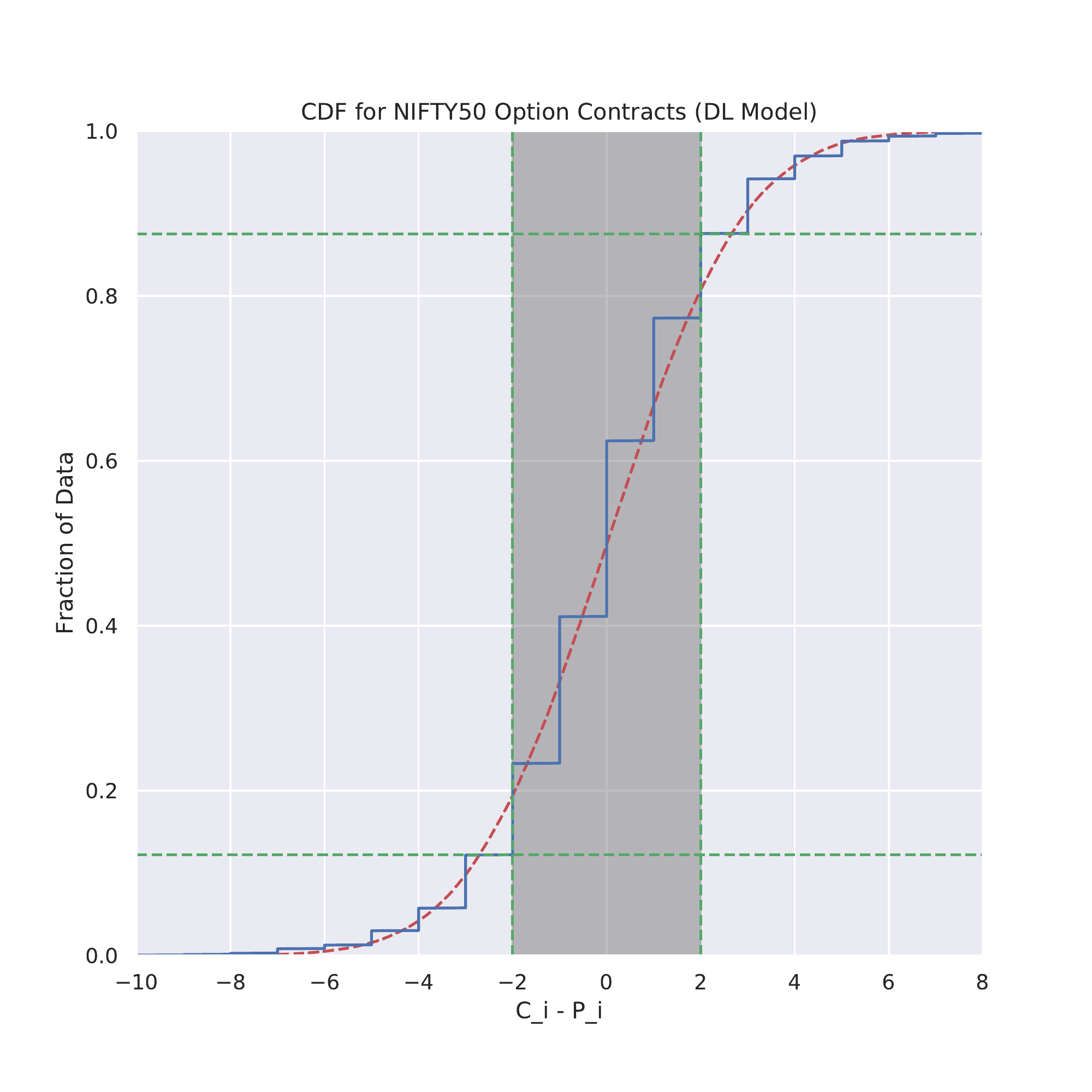}
    \caption{Empirical CDF of $C_i-P_i$ obtained using Approach I based ANN model on NIFTY50 option contracts}
    \label{fig:ErrDistn}
\end{figure}

\noindent It is evident from Table \ref{tab:Main_Reuslts} that for all cases the EM value is less than $0.19$. Loosely speaking, this implies that on an average, the predicted value of $100 \times \frac{C}{K}$ is not further than $0.19$ from the actual (refer to Equation (\ref{EM})). In other words, the difference between the actual and predicted option prices is on an average, less than $0.0019 \times K$. A more precise statement in terms of confidence interval can be made using the empirical quantiles (refer to Figure \ref{fig:ErrDistn}). The 2\% and 98\% quantiles of $C_i-P_i$ obtained using the Approach I ANN model for NIFTY$50$ data are $-5$ and $5$ respectively. This implies that the actual price bin is within 5 neighbouring bins of the predicted bin with $96\%$ probability for the test dataset. Similarly, from other quantile values we can also deduce that the actual price bin is within 2 neighbouring bins of the predicted bin with $74\%$ confidence. Figure \ref{fig:ErrDistn} illustrates this using a plot of the empirical CDF of $C_i-P_i$. Indeed, the $\rho$ metric is useful in this regard. To be more precise, the difference between the actual and predicted option price intervals is less than $\frac{2K}{1000}$ with probability $1-\rho$ (refer Equation (\ref{rho})). Thus an interval of length $\frac{5K}{1000}$ (predicted band) can be obtained from a model prediction which succeeds in containing the close price of the option (having strike price $K$) with probability $1-\rho$ (refer to Figure \ref{fig:rho_band}). We recall from Figure \ref{fig:percentage_hist} that this predicted band width is less than one tenth of the full range of option prices for the NIFTY$50$ data under consideration. \\

\begin{figure}[!h]
    \centering
    \includegraphics[width = 0.7\textwidth]{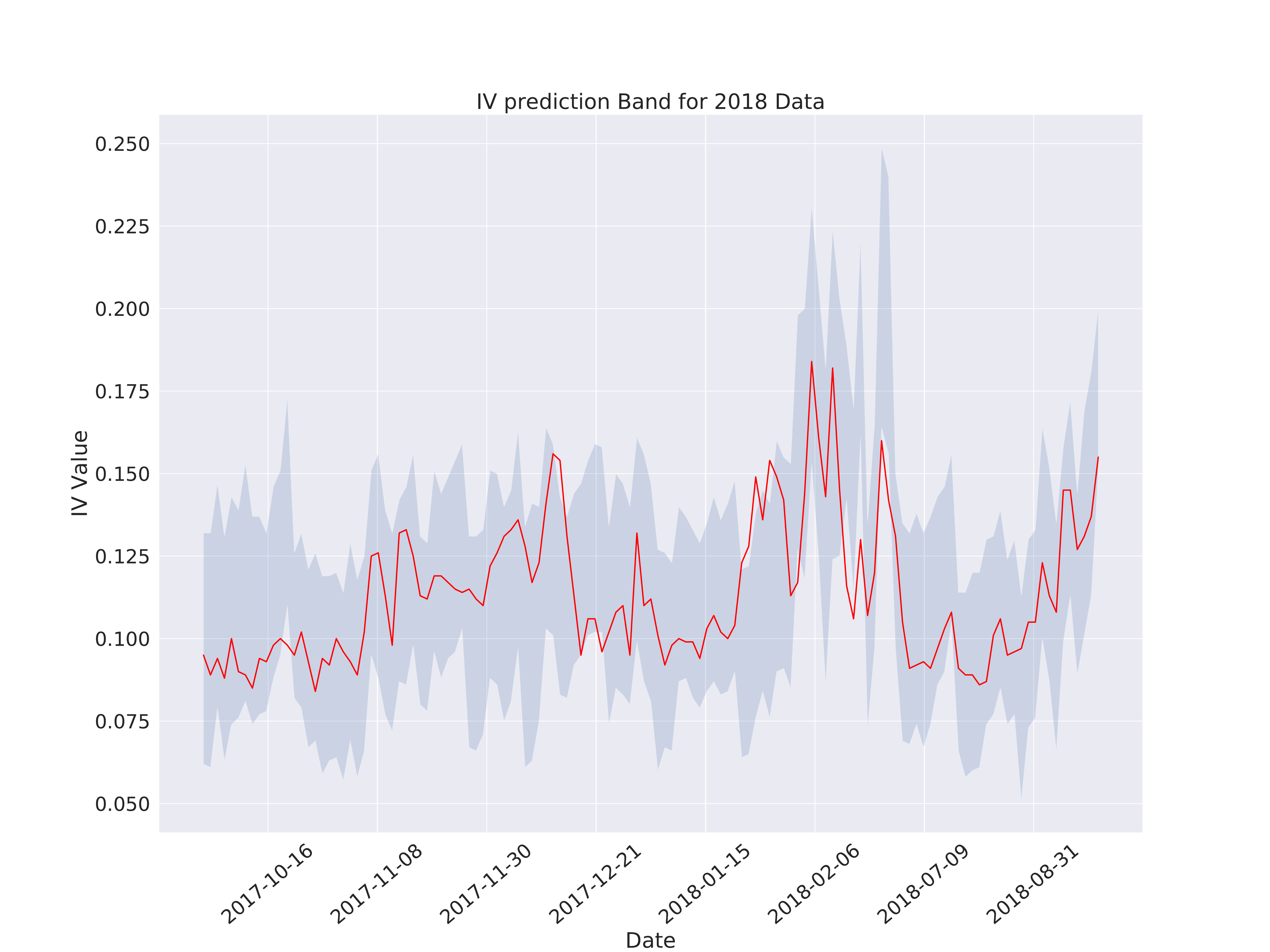}
    \caption{Average Empirical IV and the predicted IV Band, plotted for the NIFTY50 test dataset}
    \label{fig:2018_IV_bands}
\end{figure}

\noindent
We consider Approach III ANN models to further illustrate the implication of the predicted bands. For this, we first identify the upper and lower limit option prices of the band and compute the corresponding daily implied volatility values for each contract. From these values, we obtain the daily averaged predicted implied volatility band. We then compute the average market-realized implied volatility for each day using near-ATM options data and compare it with the predicted implied volatility band. A time series plot of that comparison is presented in Figure \ref{fig:2018_IV_bands}. The figure shows that for $90\%$ of the time, the market-realized implied volatility lies within the predicted band. It is not surprising that this band prediction error is only $0.10$, a value that is much lesser than the $\rho$ value for Approach III in Table \ref{tab:Main_Reuslts}. The main reason behind the observed error reduction is the presence of averaging in the computation. This indicates the possibility of building a superior hybrid model by exploiting such an averaging effect. However, we do not attempt to build such models in the present study. \\

\paragraph{\textbf{BANKNIFTY Index option data}}

Table \ref{tab:BNF_Main_Reuslts} lists out the performance of the models that were trained and tested on BANKNIFTY Index data. The data processing, feature-set generation and train-test splitting for the BANKNIFTY options dataset is done in the exact same way as for NIFTY$50$ Index option data, in accordance with the methodologies laid down in Sections \ref{3} and \ref{4}. It can clearly be seen that the values of the EM and $\rho$ are the lowest for Approach III models. The evaluation metrics for Approach III are also lower than those for the Black-Scholes formula. Using the trained models and the results shown in Table \ref{tab:BNF_Main_Reuslts}, an analysis of the results in a manner similar to what has been done for NIFTY trained models can be performed, but we avoid repetitive explanation.
\begin{table}[!h]
\centering
\begin{tabular}{@{}ccccc@{}}
\toprule
\multicolumn{5}{c}{\textit{\textbf{BANKNIFTY Contracts}}}                        \\ \toprule
& \multicolumn{2}{c}{\multirow{2}{*}{\textbf{EM}}} & \multicolumn{2}{c}{\multirow{2}{*}{\textbf{$\rho$}}} \\
 & \multicolumn{2}{c}{} & \multicolumn{2}{c}{} \\ \cline{2-5}
B-S Pricing &  \multicolumn{2}{c}{0.19} & \multicolumn{2}{c}{0.29}\\ \cline{2-5}
\textit{\textbf{Trained Models}}  & \textit{\textbf{ANN}} & \textit{\textbf{XGB}} & \textit{\textbf{ANN}}      & \textbf{XGB}     \\ \midrule
Approach I   & 0.19     & 0.21   & 0.28    &  0.32     \\
Approach II  & 0.20     & 0.21     & 0.33     &  0.33    \\
Approach III & 0.17     & 0.19     & 0.24    &  0.29    \\ \bottomrule
\end{tabular}
\caption{Model evaluation metrics for models trained and tested on BANKNIFTY options contract price data}
\label{tab:BNF_Main_Reuslts}
\end{table}\\

\noindent From the results shown in Table \ref{tab:BNF_Main_Reuslts} and Table \ref{tab:Main_Reuslts}, it is evident that Approach III ANN models perform significantly better than all other proposed models. Furthermore, they are far more accurate than what the Black-Scholes formula can prescribe. Having said so, it is also important to recall that no measures of volatility has been fed into any of the proposed models. We next present a set of experiments that show the promise of ensemble modeling.

\subsection{Ensemble Models}
\noindent The predictions of the two pricing models obtained using ANN and XGBoost for each approach can be averaged out, to obtain a new prediction. We refer to this as the prediction of a simple ensemble model. The rationale behind this approach is straightforward. It is plausible that for a particular approach, the XGBoost model learns a subset of the representation space very well, but does not learn it well enough for some other subsets. The ANN model could hypothetically learn those missed subsets of representation space better than what the XGBoost model is capable of learning. By averaging out the predictions of the models, we seek to minimize the number of subsets over which the individual models perform poorly. Averaging the model predictions allow us a way to leverage the well learnt portions of the representation space of both the models at the same time.\\

\noindent We evaluate the performance of the ensemble models by computing the EM and the $\rho$ values for the test sets. Tables \ref{tab:average_model1} and \ref{tab:average_model2} present the model evaluation metric values for the ensemble models trained and tested on NIFTY$50$ and BANKNIFTY contracts respectively. The results in Table \ref{tab:average_model1} and Table \ref{tab:average_model2} show a marked improvement in the EM values for all the approaches when compared to the results in Table \ref{tab:Main_Reuslts} and Table \ref{tab:BNF_Main_Reuslts} \\ respectively.
\begin{table}[!h]
\centering
\begin{tabular}{ccc}
\hline
\multicolumn{3}{c}{\textit{\textbf{Averaged Models :: NIFTY50}}} \\ \hline
                               & \textbf{EM}            & \textbf{$\rho$}            \\ \hline
\textit{Approach I}            & 0.16                   & 0.26                     \\
\textit{Approach II}            & 0.16                   & 0.27                      \\
\textit{Approach III}            & 0.13                   & 0.19                      \\ \hline
\end{tabular}
\caption{Model evaluation metrics for ensemble averaged models trained and tested on NIFTY50 option contracts}
\label{tab:average_model1}
\end{table}

\begin{table}[!h]
\centering
\begin{tabular}{ccc}
\hline
\multicolumn{3}{c}{\textit{\textbf{Averaged Models :: BANKNIFTY}}} \\ \hline
                               & \textbf{EM}            & \textbf{$\rho$}            \\ \hline
\textit{Approach I}            & 0.18                   & 0.29                     \\
\textit{Approach II}            & 0.18                   & 0.30                      \\
\textit{Approach III}            & 0.15                   & 0.23                      \\ \hline
\end{tabular}
\caption{Model evaluation metrics for ensemble averaged models trained and tested on BANKNIFTY option contracts}
\label{tab:average_model2}
\end{table}

\noindent
\begin{rem} It is important to note that the ``predictions" ($P$) of the ensemble model need not be an integer class label but could instead be an integer multiple of $\frac{1}{2}$. However, no change is needed in the computation scheme of the model evaluation metrics.
\end{rem}

\subsection{Models Trained with Multiple Sources} \label{5.3} Since the features and the output variable used are scale free, models trained on option price data of one asset should be able to give reasonable option price predictions corresponding to another asset provided the log return distributions are not too different from each other. This anticipation hinges on our assumption that, for a given financial market, two assets having the same return distribution should have the same option pricing mechanism. Again, there is a possibility that the prediction quality may be inferior even though the training and test datasets belong to the same asset,  as possibly the return dynamics of the underlying asset may have changed drastically. This subsection presents some experiments in this direction.\\

\begin{figure}[!h]
    \centering
    \includegraphics[width = 0.55\textwidth]{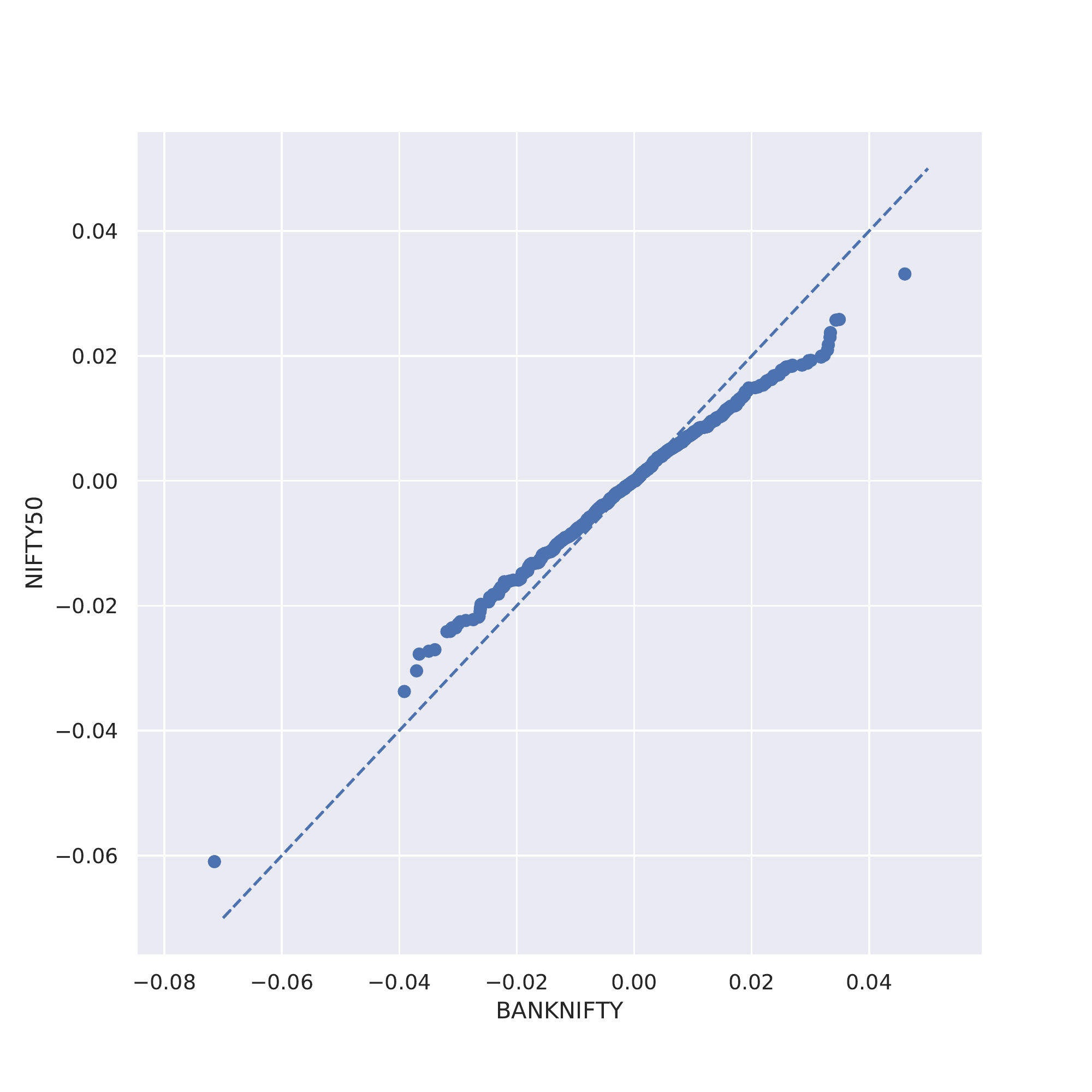}
    \caption{Q-Q plot for the log returns distributions of the ``Close" prices of the BANKNIFTY and NIFTY50 indices}
    \label{fig:QQ_plot}
\end{figure}

\noindent We first carry out an empirical investigation on the asset portability of the models. In order to do this we consider all six models trained on NIFTY$50$ option contracts, and test them with data from BANKNIFTY based contracts on non-overlapping time intervals. The results of this experiment are given in Table \ref{tab:NF(BNF)_port}. It is crucial to note that these two indices are sufficiently independent and have contract parameters with vastly different magnitudes. We present the Q-Q plot (Figure \ref{fig:QQ_plot}) of the ``Close" price log returns of the two underlying assets in order to compare their log return distributions. Figure \ref{fig:QQ_plot} shows a moderate mismatch between the return distributions of these two assets. Thus, although we do not expect the predictive performance to be equivalent to NIFTY$50$ test sets, we expect the error metric to be decently small in magnitude. Our experiment supports this anticipation.
\begin{table}[!h]
\centering
\begin{tabular}{@{}ccccc@{}}
\toprule
\multicolumn{5}{c}{\textit{\textbf{NIFTY50 models tested on BANKNIFTY}}}              \\ \toprule
& \multicolumn{2}{c}{\multirow{2}{*}{\textbf{EM}}} & \multicolumn{2}{c}{\multirow{2}{*}{\textbf{$\rho$}}} \\
& & & \\\cline{2-5}
\textit{\textbf{Trained Models}}  & \textit{\textbf{ANN}} & \textit{\textbf{XGB}} & \textit{\textbf{ANN}}      & \textbf{XGB}     \\ \midrule
Approach I         & 0.19     & 0.19     & 0.28      & 0.28    \\
Approach II        & 0.17     & 0.18     & 0.24      & 0.26    \\
Approach III       & 0.15     & 0.17     & 0.21      & 0.24    \\ \bottomrule
\end{tabular}
\caption{Model evaluation metrics for models trained on NIFTY50 contract data and tested on BANKNIFTY contracts}
\label{tab:NF(BNF)_port}
\end{table}
However, a quick comparison of our results (Table \ref{tab:NF(BNF)_port}) with Table \ref{tab:BNF_Main_Reuslts} shows that the NIFTY$50$-trained models outperform the BANKNIFTY-trained models for the BANKNIFTY test set. This gives evidence of the fact that a model trained on a different asset/source can outperform a model trained on the target asset/source.\\

\noindent The results of the above experiment encourages us to train the models using contract data from two or more number of assets/sources. In principle, this should broaden the range of features and allow the models to achieve far better generalization and predictive capability. We investigate this by training all six models (one XGB and one ANN for each of the three approaches) using the combined data of NIFTY$50$ and BANKNIFTY contracts and then perform out-of-sample tests for each asset. The EM and $\rho$ values of the respective experiments are given in Table \ref{tab:NF+BNF(NF+BNF)_port}. \\

\begin{table}[!h]
\begin{tabular}{@{}ccccccccc@{}}
\toprule
\multicolumn{9}{c}{\textit{\textbf{Experiments using models trained on combined datasets}}}   \\ \midrule
                & \multicolumn{4}{c}{\textbf{EM}}       & \multicolumn{4}{c}{\textbf{$\rho$}} \\ \cline{2-9}

\textit{\textbf{Test Dataset ::}} &
  \multicolumn{2}{c}{\textbf{NIFTY50}} &
  \multicolumn{2}{c}{\textbf{BANKNIFTY}} &
  \multicolumn{2}{c}{\textbf{NIFTY50}} &
  \multicolumn{2}{c}{\textbf{BANKNIFTY}} \\ \midrule

B-S Pricing &
  \multicolumn{2}{c}{0.19} &
  \multicolumn{2}{c}{0.19} &
  \multicolumn{2}{c}{0.29} &
  \multicolumn{2}{c}{0.29} \\ \midrule

\textit{\textbf{Experiment Type}} &
  \textit{\textbf{ANN}} &
  \textit{\textbf{XGB}} &
  \textit{\textbf{ANN}} &
  \textit{\textbf{XGB}} &
  \textit{\textbf{ANN}} &
  \textit{\textbf{XGB}} &
  \textit{\textbf{ANN}} &
  \textit{\textbf{XGB}} \\ \midrule
Approach I      & 0.17    & 0.17    & 0.18    & 0.19    & 0.24    & 0.24    & 0.25   & 0.25   \\
Approach II     & 0.17    & 0.18    & 0.19    & 0.19    & 0.25    & 0.28    & 0.28   & 0.28   \\
Approach III    & 0.14    & 0.16    & 0.16    & 0.17    & 0.17    & 0.22    & 0.23   & 0.23   \\ \bottomrule
\end{tabular}
\caption{Model evaluation metrics for models trained on both NIFTY50 and BANKNIFTY contract data}
\label{tab:NF+BNF(NF+BNF)_port}
\end{table}

\begin{figure}[!h]
    \centering
    \includegraphics[width = 0.75\textwidth]{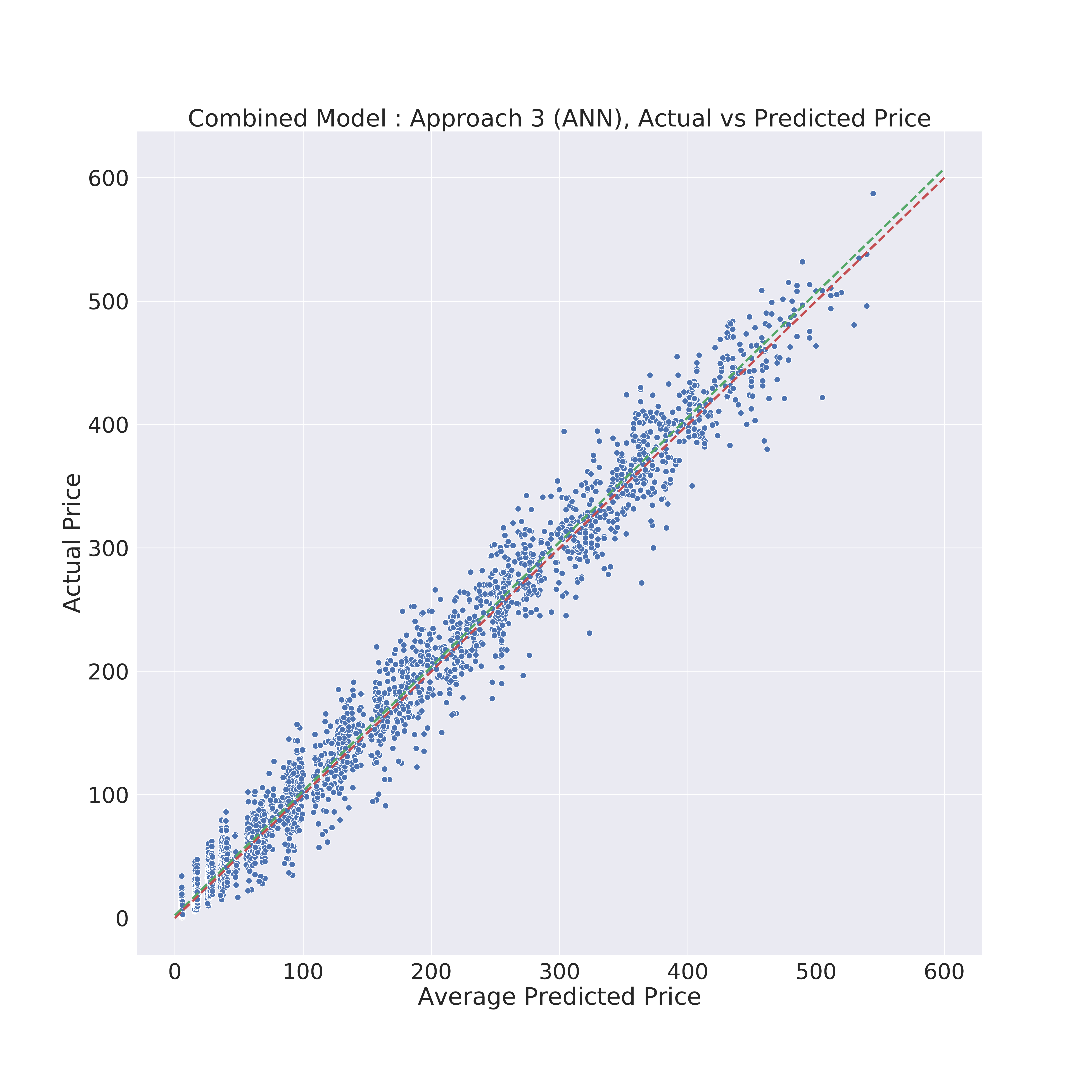}
    \caption{Actual vs `Predicted' Price (obtained using the Approach III ANN model trained on both NIFTY$50$ and BANKNIFTY contract data)}
    \label{fig:Actual_vs_Predicted_Combined_Approach_3_ANN}
\end{figure}

\noindent
A comparison of the metrics given in Table \ref{tab:NF+BNF(NF+BNF)_port} with those in Tables \ref{tab:Main_Reuslts} and \ref{tab:BNF_Main_Reuslts} clearly shows that combined-trained models have better option pricing capabilities than the models trained on the respective assets individually. Each of the combined-trained models also outperform the price prescription of the Black-Scholes formula. The performance of the option price prediction can be better perceived using a scatter plot of the actual and predicted option prices, which we present in Figure \ref{fig:Actual_vs_Predicted_Combined_Approach_3_ANN}. Since the proposed models predict a bin, in order to plot the graph (Figure \ref{fig:Actual_vs_Predicted_Combined_Approach_3_ANN}) we take the mid point of the predicted bin to get a single predicted price. The prices obtained using the mid point of the bins are plotted along the horizontal axis and the closed price from test data is plotted along the vertical axis in the scatter plot. The scatter plot shown in Figure \ref{fig:Actual_vs_Predicted_Combined_Approach_3_ANN} is constructed using the predictions given by the Approach III ANN model (trained using combined data). In principle, such scatter plots can be constructed for all the proposed models.

\begin{rem} To the plot \ref{fig:Actual_vs_Predicted_Combined_Approach_3_ANN}, we add the identity line $y=x$ (red dashed line) and the orthogonal regression line (green dashed line). The proximity of these two lines in the plot validates the absence of bias in the model. Since presence of a bias corresponds to over-pricing or under-pricing, the proximity of the regression line to the identity line cross-validates fair-pricing by the trained model. We propose this analysis as a good method to empirically verify arbitrage-free pricing of a data-driven model.
\end{rem}

\noindent The success of the above experiment warrants an in-depth explanation.
In the next section, we use the concept of domain adaptation for designing a methodology that provides a deeper understanding of the combined-training effect.

\section{Introspection of Combined trained Models}\label{6}
\noindent
This section brings to fore an interesting application of the models constructed using Approach I in the Sections \ref{5.2} and \ref{5.3} respectively. We test the pre-trained models (trained using Approach I feature sets) with simulated Black-Scholes option price data. A family of such tests is conducted by varying the volatility parameter in the Geometric Brownian motion that is used to generate the simulated asset price time series; these time series datasets are then augmented with the option prices prescribed by the Black-Scholes formula. We recall from Section \ref{4.2} that Approach I based models use Order Statistics of the log returns of the underlying asset's daily close prices as their primary inputs. Thus by considering the simulated time series data as ``Close" prices, Approach I can directly be used to generate the feature sets. But Approach II or Approach III cannot be used directly, as they use ``Open", ``High" and ``Low" time series' along with the ``Close" time series to generate the features, and simulating the corresponding ``Open", ``High" and ``Low" time series' is not straightforward. Hence we only use Approach I based models for the experiments described in this section. \\

\noindent
We simulate Geometric Brownian motion with the drift parameter set at $\mu = 0.1$ and vary the volatility parameter from $1\%$ to $20\%$ using an increment of $1\%$. Daily data is simulated for each value of the volatility parameter, such that we obtain a test set that represents a trading session of $500$ days. This test data is augmented with the price of several near-ATM option contracts (with values of time to maturity $\in [10, 25, 40]$) using the Black Scholes formula. We then find the prediction error of the models for each variant of the test data and plot them against the volatility parameter. We do this using XGBoost/ANN models trained on NIFTY$50$, BANKNIFTY and the combined dataset respectively. The purpose of this exercise is to explain the results detailed in Section \ref{5.3}. It has not been done to judge the performance of the trained models on data derived from theoretical models (as option contract prices obtained using theoretical models involve certain mathematical assumptions that renders the pricing obtained dissonant from reality).  \\

\begin{figure}[h]
    \centering
    \includegraphics[width = 0.55\textwidth]{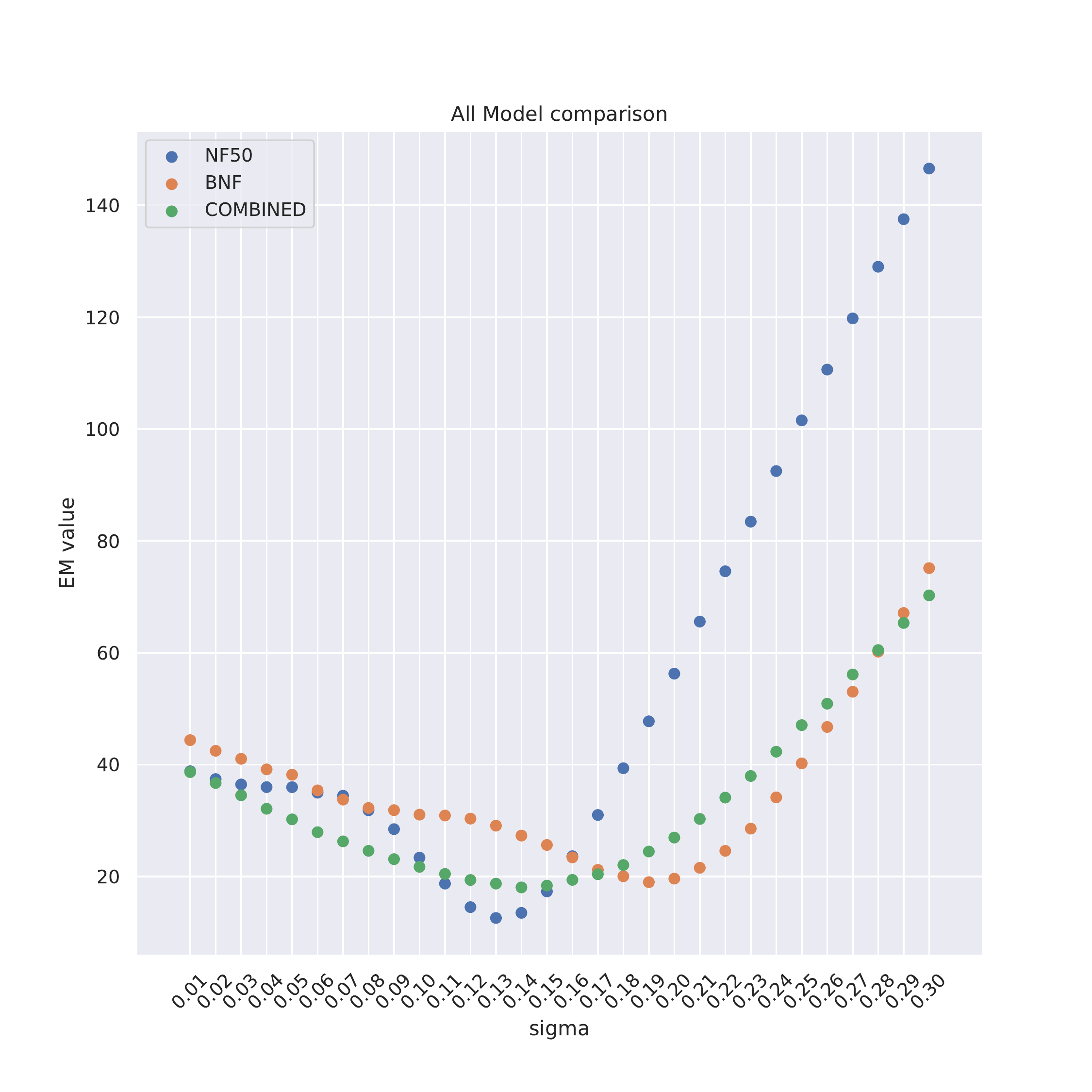}
    \caption{EM vs $\sigma$ curve for ANN single- and combined- trained models in Approach I}
    \label{fig:TheoEM_gBm_ANN}
\end{figure}

\begin{figure}[h]
    \centering
    \includegraphics[width = 0.55\textwidth]{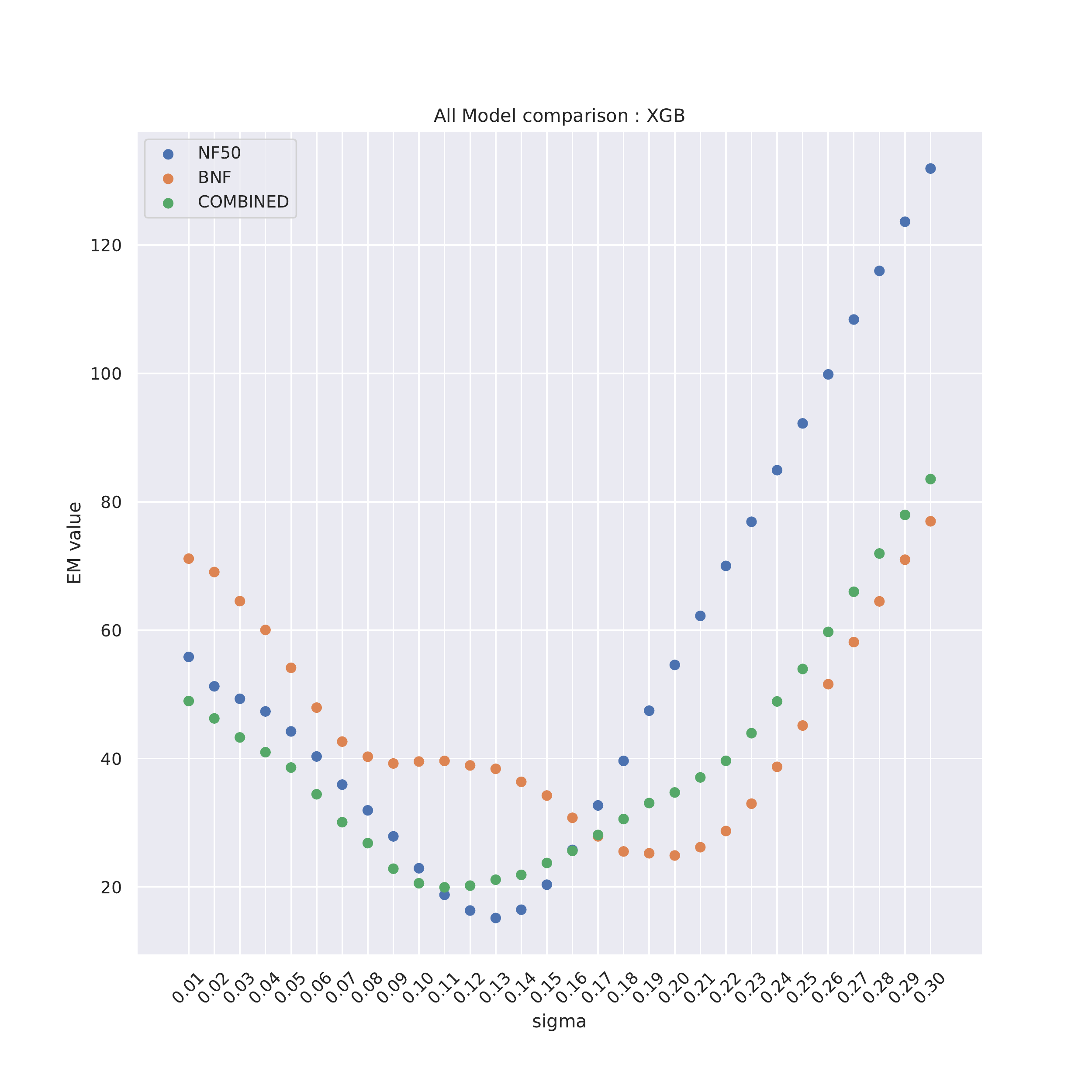}
    \caption{EM vs $\sigma$ curve for XGBoost single- and combined- trained models in Approach I}
    \label{fig:TheoEM_gBm_XGB}
\end{figure}

\noindent
We recall from Section \ref{5.3}, that the success of a model on a test dataset depends on the proximity between the return distributions of the test set and the model's training set. Hence, it is natural to expect that the quality of the trained model's predictions will vary when the model is tested with the range of simulated datasets with varying volatilities. We anticipate that the plot of the error metric (obtained for each of the test sets) against the value of the test set's volatility parameter would result in a $V$ shaped curve as the error metric would be large for a test set having a higher mismatch with the training set in terms of the return distributions. This is clearly observed in Figure \ref{fig:TheoEM_gBm_ANN} and \ref{fig:TheoEM_gBm_XGB}. \\

\noindent The minimizing volatility values of EM provide a class of theoretical asset dynamics whose option prices are best predicted by the trained model. We call it the ``Error Minimizing Volatility" or EMV of a given option price dataset corresponding to the learning model. For example from Figures \ref{fig:TheoEM_gBm_ANN} and \ref{fig:TheoEM_gBm_XGB}, it can be observed that the EMV of NIFTY$50$ data for the ANN as well as the XGBoost model is $0.13$. On the other hand the EMV of BANKNIFTY data for the ANN and XGBoost models are $0.19$ and $0.20$ respectively. Examining Table \ref{tab:iv_table} shows us that the values of EMV obtained are not mere coincidences, but are in fact related to the training dataset used. Indeed the EMV is significantly close to the volatility parameter values of the training dataset. \\
\begin{table}[h]
\centering
\begin{tabular}{@{}ccccc@{}}
\toprule
\multirow{2}{*}{} & \multicolumn{3}{c}{\textbf{Implied Volatility}} & \multirow{2}{*}{\textbf{\begin{tabular}[c]{@{}c@{}}Historical Volatility\end{tabular}}} \\ \cline{2-4}
                            & \textit{Mean} & \textit{Median} & \multicolumn{1}{c}{\textit{Mode}} &       \\ \midrule
\textit{\textit{NIFTY50}}     & 0.137         & 0.135           & 0.14                               & 0.141 \\
\textit{\textit{BANKNIFTY}} & 0.186         & 0.189           & 0.2                                & 0.191 \\ \bottomrule
\end{tabular}
\caption{IV and Historical Volatility values for NIFTY50 and BANKNIFTY indices}
\label{tab:iv_table}
\end{table}

\noindent From Figures \ref{fig:TheoEM_gBm_ANN} and \ref{fig:TheoEM_gBm_XGB} it is evident that the EM plot obtained for the combined-trained models give lower and flatter $V$ shape curves. This implies that models trained on the combined dataset result in lower EM values for a wide range of test sets having varying $\sigma$ values. This hints at the possibility of domain adaptability of predictive models trained on datasets derived from multiple assets/sources. It also hints at the existence of a common representation space for datasets with similar log return distributions. Such an application of domain adaptability can be a very powerful method, as it could potentially aid research in areas where data is scarce. \\

\section{Model performance on 2019-2020 data}\label{7}

\begin{figure}[!h]
    \centering
    \includegraphics[width = 0.55\textwidth]{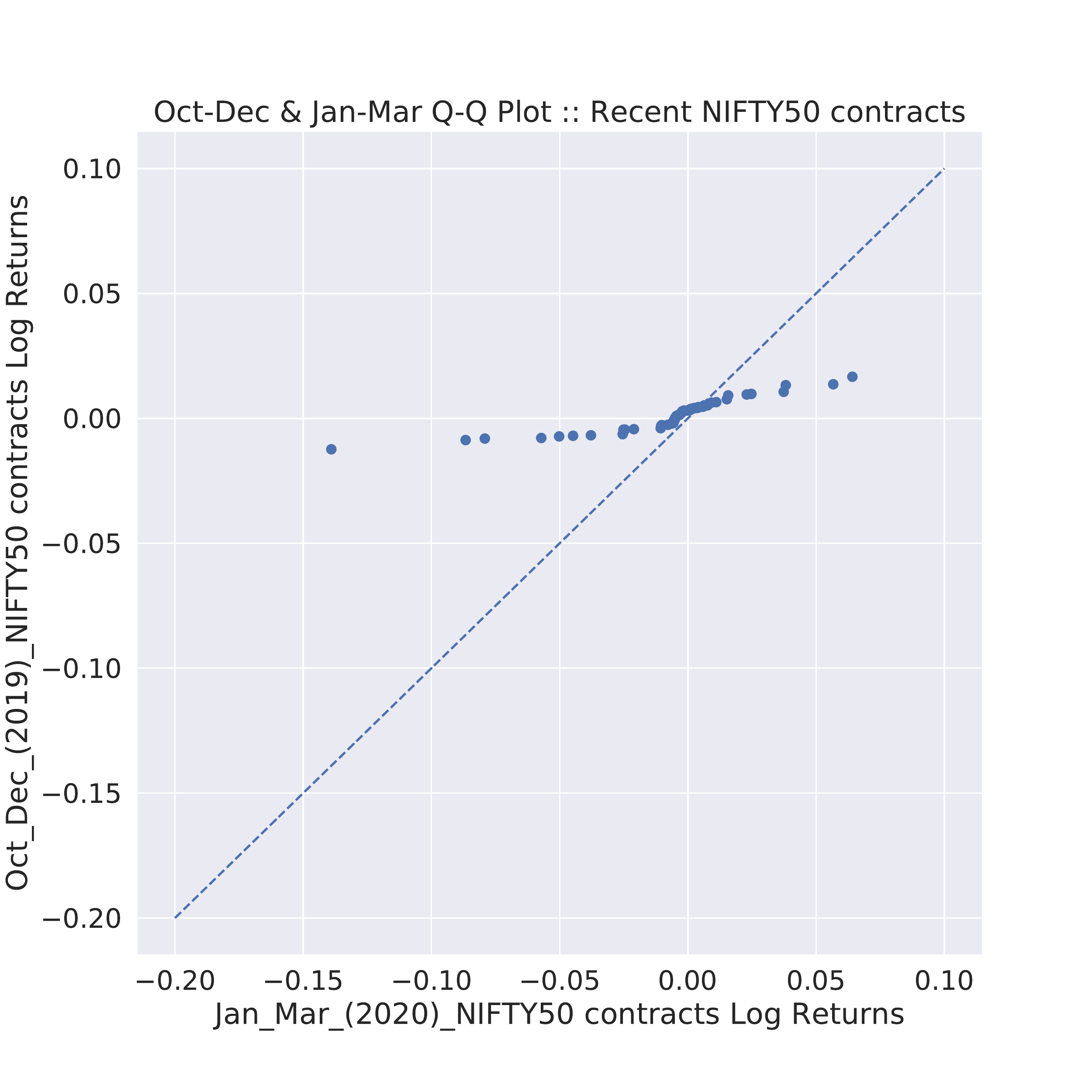}
    \caption{Q-Q plot of (Oct 2019 - Dec 2019) and (Jan 2020 - Mar 2020) datasets}
    \label{fig:recent_data_qq_plot}
\end{figure}

\noindent  During the period from January 2020 to April 2020 of the COVID-19 pandemic, the dynamics of the NIFTY$50$ Index were radically different from its usual dynamics. A Q-Q plot comparison of the log return distributions of the NIFTY$50$ Index during the periods Oct'19-Dec'19 and Jan'20-Mar'20 is shown in Figure \ref{fig:recent_data_qq_plot}. \\

\noindent
It is evident from the Q-Q plot that there seems to be almost no match between the price dynamics of these two time intervals. Therefore for option contracts based on the NIFTY$50$ index, we cannot expect the models trained on $2015-2017$ data to perform well on $2019-2020$ data. Table \ref{tab:2019_EM_test} presents the values of the performance metrics, for when the pre-trained Approach III models (constructed in sections \ref{5.2} and \ref{5.3}) are tested on $2019-2020$ data for the NIFTY$50$ Index. We consciously make the choice to use models constructed using Approach III as the benchmark to test the $2019-2020$ dataset, as these models have given us the best predictive capability. \\

\begin{table}[!h]
\begin{tabular}{@{}ccccc@{}}
\toprule
\multicolumn{5}{c}{\textit{\textbf{Testing Recent Data (Approach III)}}} \\ \midrule
                          & \multicolumn{2}{c}{\textbf{EM}} & \multicolumn{2}{c}{\textbf{$\rho$}} \\ \cline{2-5}
\textit{B-S Pricing}      & \multicolumn{2}{c}{0.30}        & \multicolumn{2}{c}{0.45}            \\ \midrule
\textbf{Train Dataset}          & \textbf{ANN}   & \textbf{XGB}   & \textbf{ANN}     & \textbf{XGB}     \\ \midrule
\textit{NIFTY50}          & 0.23           & 0.24           & 0.27             & 0.29             \\
\textit{Combined Dataset} & 0.21           & 0.22           & 0.27             & 0.28             \\ \bottomrule
\end{tabular}
\caption{Model evaluation metric for models trained on 2015-2017 NIFTY$50$ contract data but tested on $2019$ NIFTY50 contract data}
\label{tab:2019_EM_test}
\end{table}

\noindent
Table \ref{tab:2019_EM_test} makes it evident that the error in predicting option prices for $2019-2020$ NIFTY$50$ test data is significantly larger in comparison to the prediction error for $2017-2018$ NIFTY$50$ test data as in Table \ref{tab:Main_Reuslts}. It must be noted that performance of the models on the recent data is far better than what the Black Scholes formula prescribes. The large value of the evaluation metrics for the Black-Scholes pricing formula implies a large gap between the historical and implied volatilities. This is typically observed when drastic changes occur in a financial market. We also observe a significant improvement in case of the combined-trained models as compared to the individually-trained NIFTY50 models. This reaffirms the power of combined training. \\

\begin{figure}[h]
    \centering
    \includegraphics[width = \textwidth]{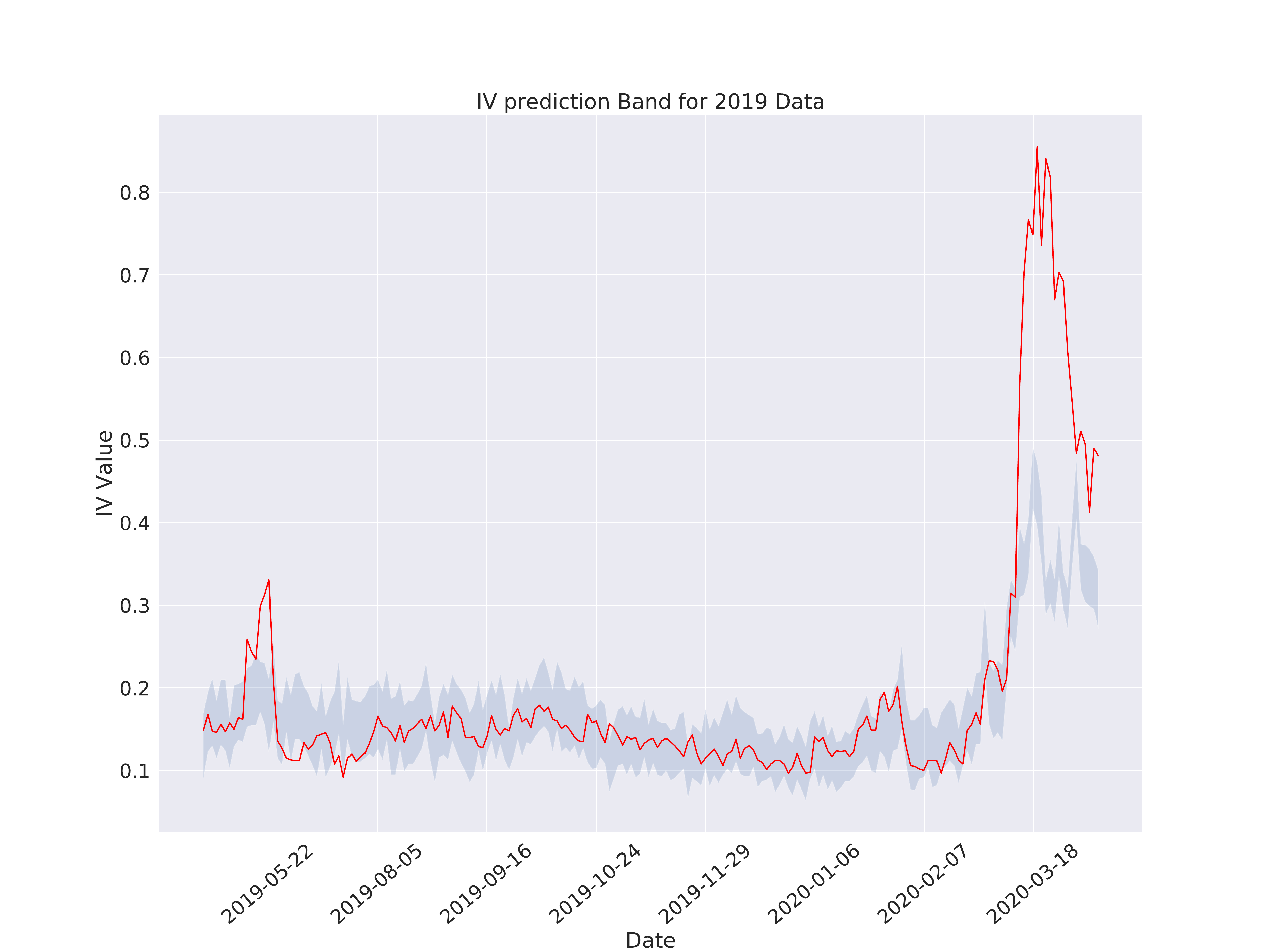}
    \caption{Average IV and the IV band for 2019 NIFTY50 index data}
    \label{fig:2019_IV_bands}
\end{figure}

\noindent
In addition to the above experiments, we plot the empirical IV and the predicted IV band in Figure \ref{fig:2019_IV_bands} in a manner similar to the plot reported in Figure \ref{fig:2018_IV_bands}. The band prediction error for the $2019-2020$ dataset (Figure \ref{fig:2019_IV_bands}) is $25\%$, which is lesser than the value of $\rho$ observed in Table \ref{tab:2019_EM_test}. Figure \ref{fig:2019_IV_bands} helps us identify regions in the test dataset where the model does not perform well. It is observed that when the implied volatility of the underlying asset changes sharply, the prediction bands deviate from the actual values. These abrupt changes are usually caused by rapid changes in the market sentiment (in this case due to the COVID-$19$ pandemic); an aspect that is not represented in the data used to train the models.

\section{Conclusion}\label{8}
\noindent In this paper, we present three data-driven approaches to build option pricing models using supervised learning algorithms. These approaches are illustrated for two different assets/sources (NIFTY$50$ and BANKNIFTY), and we use two different learning algorithms to build a range of models. Upon evaluating the performance of the models on out-of-sample data, it was seen that Approach I and II based models performed better than the Black-Scholes option pricing formula in most cases, while the Approach III based models performed significantly better than all comparative models. Since Approach III uses features derived from the historical option price data that are not present in the Approach I and II based feature sets, the performance improvement clearly indicates the vitality of including such information. The results also highlight the superior performance of ANN-based models in comparison to the XGBoost-based models. In this paper, we have also attempted to build “averaging ensemble” models for each data source; the results of which clearly shows an unprecedented level of accuracy in pricing option contracts. Lastly, we have investigated the effect of multi-asset combined training for each of the proposed approaches. It was observed that the multi-asset trained models gave us a significant improvement in the prediction quality when compared to single-asset trained models. We have further examined this performance enhancement by using the concept of domain adaptation.\\

\noindent The success of the multi-asset trained models makes us optimistic about the viability of building a non-asset-specific data-driven option pricing model. Such a model—once trained on data from multiple assets belonging to a particular financial market—would be capable of predicting the fair price of any European-style call option on any asset belonging to the same financial market with a high degree of precision. However, in our paper, we have examined the combined-training effect using only two assets/sources. Extensive experimentation is required to determine the limitations and the scope of such non-asset-specific models. Readers may refer to \cite{SC} which reports a similar extensive experiment to study some other universal non-asset-specific relations captured by a deep learning model. Further research to develop and validate the existence of such models has been planned by the authors. The codes used in this study can be made available on request.
 \\


\section*{Acknowledgement} \noindent We are grateful to Arkaprava Sinha and Prof. Amit Mitra (IIT- Kanpur) for some useful discussions.

\section*{Appendix}
\noindent
We report the hyperparameter values used to train the models described in this manuscript (refer sections \ref{2} and \ref{4}).

\begin{table}[!h]
\centering
\begin{tabular}{@{}cc@{}}
\toprule
\textbf{\begin{tabular}[c]{@{}c@{}}Hyperparameter \\ Name\end{tabular}} & \textbf{Value Set} \\ \toprule
n\_estimators                                                           & 100                \\
max\_depth                                                              & 3                  \\
learning\_rate                                                          & 0.3                \\ \bottomrule
\end{tabular}
\caption{Hyperparameter Values for XGBoost}
\label{tab:hyperparams_xgb}
\end{table}

\begin{table}[!h]
\centering
\begin{tabular}{@{}cc@{}}
\toprule
\textbf{\begin{tabular}[c]{@{}c@{}}Hyperparameter \\ Name\end{tabular}} & \textbf{Value Set} \\ \toprule
batch\_size                                                             & 32                 \\
learning\_rate                                                          & 0.00012            \\ \bottomrule
\end{tabular}
\caption{Hyperparameter Values for ANN}
\label{tab:hyperparams_ann}
\end{table}


\bibliographystyle{bibft}\it
\bibliography{bibfile}

\appendix

\end{document}